\documentclass[12pt]{article}
\usepackage[a4paper, margin=1in]{geometry}
\usepackage{graphicx} 
\usepackage{amscd}
\usepackage{verbatim}
\usepackage{hyperref}
\usepackage{amssymb}
\usepackage{amsthm}
\usepackage{amsmath}
\usepackage{subcaption}
\usepackage{cancel}
\usepackage{tipa}
\usepackage{tabularx}
\usepackage{mathtools}
\usepackage{appendix}

\DeclarePairedDelimiterX\braket[2]{\langle}{\rangle}{#1 \delimsize\vert #2}

\usepackage{graphicx} 
\def\be{\begin{equation}}
\def\ee{\end{equation}}

\def\ld{\lambda}
\def\tb{\textbf}
\def\dt{\delta}
\def\w{\wedge}

\def\ep{\epsilon}
\def\sg{\Sigma}
\def\bds{\boldsymbol}
\def\th{\Theta}
\def\p{\phi}

\def\pd{\partial}
\def\dl{\frac{d}{d\ld}}

\def\eqv{:=}
\title{The BEF symplectic form: A Lagrangian perspective}
\author{Mohd Ali, Georg Stettinger}
\date{March 2026}

\begin{document}
\begin{center}

\vspace{24pt}
{\Large \bf The BEF Symplectic Form: A Lagrangian Perspective}

\vspace{18pt}

{\bf Mohd Ali$^{1}$\footnote{mohd.ali@icts.res.in}, 
\bf Georg Stettinger$^{1}$\footnote{georg.stettinger@icts.res.in}}

\vspace{12pt}

{\it $^{1}$International Centre for Theoretical Sciences (ICTS),\\
Bengaluru, India 560089}

\vspace{18pt}

\end{center}

\begin{abstract}
In \cite{Bernardes:2025uzg}, Bernardes, Erler and F\i rat proposed a novel, elegant expression for the symplectic form on phase space applicable to non-local theories. We show that this BEF symplectic structure can be derived directly from an $L_\infty$-Lagrangian by following the covariant phase space approach. Moreover, we establish a precise relation between the BEF symplectic structure and the Barnich--Brandt symplectic form for general finite-derivative theories. In particular, we prove that for theories with second-order equations of motion, the BEF symplectic structure coincides with the Barnich--Brandt construction, thereby explaining the emergence of the canonical corner term in general relativity within the BEF approach.

We further argue that the BEF symplectic structure naturally encodes information about generic corner terms and some information about boundary conditions. In addition, we develop a general expression for the Hamiltonian in theories in $L_\infty$- form and present several explicit examples illustrating the construction.
\end{abstract}
\newpage
\tableofcontents
\newpage
\section{Introduction}
In physics, phase space is a powerful concept that geometrizes the information about states, dynamics, and conserved charges \cite{Arnold2007}. Although, a priori, it is not obvious that a phase space can be defined for all systems, most physically interesting systems do admit such a structure.
\\
In classical mechanics, the canonical way of defining phase space is as the space of initial data for the equations governing the dynamics. However, this requires specifying a time slice on which the data are defined \cite{Arnold2007,Crnkovic:1986ex}. This perspective has two drawbacks: First, the choice of time slices appears to break covariance. For example, in relativistic classical field theory, choosing a particular time coordinate breaks Lorentz covariance and treats time as a special direction. The second problem arises for theories such as string theory, or more generally for theories that are nonlocal or involve infinitely many derivatives in the Lagrangian. In such cases, the canonical definition of phase space becomes problematic, since the theory may be nonlocal in time and may not admit a well-defined Cauchy initial value problem.

An alternative approach is to define phase space as the space of solutions to the equations of motion of the theory \cite{Crnkovic:1986ex}. With this definition, covariance is manifest, and we therefore refer to it as the covariant phase space \cite{Crnkovic:1986ex,LeeWald1990,Iyer:1994ys,Wald:1990mme, Golshani:2026lvc, Adami:2024gdx}. Moreover, this definition is consistent with the canonical one whenever the equations of motion admit a well-posed initial value problem. In that case, the space of inequivalent initial data sets is in one-to-one correspondence with the space of solutions. In this way, we bypass the first problem.

Furthermore, one can define a symplectic structure by introducing a closed and non-degenerate two-form on the space of solutions. This symplectic structure can be constructed in a covariant manner: It is obtained by integrating a conserved $(d-1)$-form over a Cauchy surface. Importantly, although the construction involves a choice of hypersurface, the resulting symplectic structure is independent of this choice since it is closed.

The second problem, however, is more subtle. For nonlocal theories, it is not immediately clear how to define a symplectic structure, since the usual notion of locality in time, which underlies the standard construction may fail. In particular, the absence of a well-defined local Cauchy problem makes it unclear how to localize degrees of freedom. Non-local deformations in general quantum field theories and string field theories were studied in \cite{Chiaffrino:2021uyd, Schnabl:2023dbv, Schnabl:2024fdx, Stettinger:2024hkp, Erbin:2023hcs, Maccaferri:2024puc, Munster:2012gy, Moeller:2010mh}. There has also been a lot of work on the initial value problem \cite{Erbin:2021hkf} as well as the boundary problem \cite{Firat:2024kxq, Stettinger:2024uus, Maccaferri:2025orz, Maccaferri:2025onc} in string field theory and related toy models, but still open questions remain.

Nevertheless, the difficulty can be addressed by introducing an auxiliary function (“sigmoid”-type function) that effectively creates a fiducial boundary and helps us to define a sort of "smeared" Cauchy surface. Now, even non-local theories exhibit a form of localization of their degrees of freedom to the region where the sigmoid is changing. This allows one to define a symplectic structure indirectly, replacing the Cauchy slice with the sigmoid.

This idea has appeared in earlier works, notably in that of Witten \cite{Witten1986}, and more recently in that of Bernardes, Erler and F\i rat (BEF) \cite{Bernardes:2025uzg, Bernardes:2025zzu, Bernardes:2025zkj, Bernardes:2026egp, BEFcharges, Bernardes:2026rsk, Bernardes:2026kpx}, where a concrete expression for the  symplectic structure in non-local theories is proposed. The BEF-proposal is formulated for any theory that admits a Lagrangian description in terms of an \(L_\infty\)-algebra. Such algebras originated in the description of the dynamics of classical string field theory \cite{Sen:2024nfd}, but also naturally arise in the gauge algebra of higher spin theories \cite{Berends:1984rq} . In fact, any Lagrangian field theory can be written in this form, for more details see \cite{Hohm:2017pnh, Jurco:2018sby}.\\
In this paper, we present the derivation of the BEF symplectic structure starting from an arbitrary \(L_\infty\)-Lagrangian and an appropriate sigmoid function. We then follow the covariant phase space prescription, as developed for finite-derivative theories, to derive the symplectic structure. Moreover, we establish a precise relation between the BEF symplectic structure and the Barnich-Brandt symplectic form for general finite-derivative theories \cite{Barnich:2001jy} in presence of a spatial boundary. In particular, we show that for theories with second-order equations of motion, the BEF symplectic current coincides with the Barnich-Brandt current. This explains the emergence of the canonical corner term in general relativity within the BEF framework.

The relation between the BEF and Barnich-Brandt constructions is not accidental. Since both are defined using the equations of motion, one expects them to be closely related. This observation also implies that, like the Barnich-Brandt symplectic form, the BEF symplectic structure is invariant under ambiguities in the choice of Lagrangian and presymplectic potential. Therefore, it is natural to interpret the BEF construction as an infinite-derivative generalization of the Barnich-Brandt homotopy construction of an invariant symplectic form.

In addition, we develop an expression for the Hamiltonian for \(L_\infty\)-theories and discuss its properties. 
\\
The outline of the paper is as follows: In Section II, we review the definition and properties of \(L_\infty\)-algebras. We then review how the symplectic structure is defined in the covariant phase space formalism, followed by a review of the BEF proposal. 

In Section III, we derive the BEF symplectic structure from an \(L_\infty\)-Lagrangian and demonstrate its relation to the covariant phase space formalism. In Section IV, we establish the consistency of the BEF symplectic form by verifying its closedness, gauge invariance, and independence of the sigmoid function in the presence of spatial boundaries. We then briefly review the Barnich-Brandt symplectic form and show its equivalence to the BEF symplectic form for second-order theories. Furthermore, we obtain a relation between the two in general finite-derivative theories. 
In Section V, we construct the Hamiltonian corresponding to an arbitrary vector field for any \(L_\infty\)-Lagrangian. We also identify the Noether current and the presymplectic potential. In Section VI, we explicitly compute the symplectic structure and Hamiltonian for Maxwell theory, a higher-derivative scalar theory, and the non-covariant Schrödinger theory. Finally, in Section VII, we present our conclusions and outlook.
\section{Review}
In this section, we briefly review \(L_\infty\)-algebras and their properties, the covariant phase space formalism and the BEF proposal.
\subsection{$L_\infty$-algebras}
Since the BEF symplectic form is formulated in the framework of $L_\infty$-algebras \cite{Bernardes:2025uzg},  let us first review some basic facts. Although its origin is closed string field theory \cite{Zwiebach:1992ie},  it has been shown that basically any field theory can be formulated using $L_\infty$-algebras \cite{Hohm:2017pnh}. Their full power is unleashed only in the presence of gauge symmetries though.

Given a graded vector space $V$ that will eventually contain the space of field configurations, define totally graded symmetric products of grade one and odd parity $L_n: V^{\otimes n}  \to  V $. They will be referred to as \textsl{higher products} and should fulfill
\be\label{L infinity symmetry}
L_n(\phi_1,...,\phi_i,\phi_{i+1},...,\phi_n) = (-1)^{|\phi_i||\phi_{i+1}|}L_n(\phi_1,...,\phi_{i+1},\phi_i,...,\phi_n) 
\ee
for all $i$, where $|\phi_i| $ denotes the Grassmann parity of $\p_i$ and 
\be\label{L inf def}
\sum_{{j_k,i_{n-k}}} \epsilon(k,n-k)L_{n-k+1}(\p_{i_1},...,\p_{i_{n-k}},L_k(\p_{j_1},...,\p_{j_k}))=0
\ee
for all $n$. Here, the sum runs over all inequivalent splittings of the indices into two groups with $k$ and $n-k$ elements and $\epsilon(k,n-k)$ denotes the Grassmann sign which is picked up. A family of $L_n$ obeying those two axioms is called an \textsl{$L_\infty$-algebra}. To define an action, there is one more algebraic ingredient needed, namely a graded antisymmetric bilinear form $\omega$:
\be\label{antisym omega}
\omega(\p_1,\p_2)=-(-1)^{|\phi_1||\phi_2|}\omega(\p_2,\p_1).
\ee
We demand $\omega$ to be non-degenerate and the \textsl{cyclicity conditions} for the higher products
\be\label{cyclicity}
\omega(\p_1,L_n(\p_2,...,\p_{n+1}))=-(-1)^{|\phi_1|}\omega(L_n(\p_1,...,\p_n),\p_{n+1}).
\ee
to hold for all $n$ \footnote{We will see that in the presence of a spatial boundary, extra terms arise in the cyclicity relations.}. Together, the $L_n$ and $\omega$ form a \textsl{cyclic} $L_\infty$-algebra. 
\\
Consider now the action 
\be\label{L inf action} 
S=-\sum_{n=1}^\infty \frac{1}{(n+1)!}\omega(\p,L_n(\p,...,\p)).
\ee
Varying and using \ref{cyclicity}
yields the equations of motion 
\be\label{eom gen}
\sum_{n=1}^\infty \frac{1}{n!}L_n(\p,...,\p)=0.
\ee
The operator $L_1$ is typically denoted by $Q$ and corresponds to the kinetic operator of the theory in question. From \ref{L inf def} we can deduce that it is nilpotent and obeys the graded Leibniz rule with respect to $L_2$. 
Let us now see how gauge transformations are encoded in this formalism: Define the operator
\be\label{Q_phi def}
Q_\p\Lambda=Q\Lambda+\sum_{n=2}^\infty\frac{1}{(n-1)!}L_n(\Lambda,\p,...,\p)
\ee
for an arbitrary gauge parameter $\Lambda$. One can show by only using the $L_\infty$-axioms as well as cyclicity that the action is invariant under transformations of the form $\delta\p=Q_\p\Lambda$. Moreover, it is easy to see that $Q_\p$ is graded cyclic with respect to $\omega$ and squares to zero if $\p$ obeys the equations of motion \ref{eom gen}. In fact, $Q_\p$ implements the linearized equations of motion around the solution $\p$, hence 
\be\label{Q_phi kills delta phi}
Q_\p \delta\p=0
\ee
for $\p$ being on-shell. For more information on $L_\infty$-actions see \cite{Hohm:2017pnh, Vosmera:2020jyw, Sen:2024nfd, Sen:2016qap}.

Let us see from the simple example of a free scalar field (see for instance \cite{Okawa:2022sjf, Konosu:2023psc, Konosu:2024zrq, Konosu:2025bnz})  how to apply the formalism in practice: Consider the action
\be
S=\frac{1}{2}\int d^dx\   \partial_\mu\p\partial^\mu\p.
\ee
First, we use partial integration to bring it into the form 
\be\label{free sclalar L inf}
S=-\frac{1}{2}\int d^dx\   \p\square \p.
\ee
Note that this step may produce a non-trivial boundary term. This is true in general, the standard form and the $L_\infty$-form of a classical field theory differ by boundary terms. From \ref{free sclalar L inf} we can read off 
\be\label{free scalar ex}
\omega(\p_1,\p_2)=\int d^dx\, \p_1  \p_2 \text{        and        } Q=\square
\ee
with all higher products vanishing. However, we are not finished yet, \ref{L inf def} are not satisfied since $\square^2\ne0$. To repair that, we need a careful definition of our vector space: Let us denote the vector space of classical field configurations by $\mathcal{F}$. Define 
\be\label{def V}
V=\mathcal{F}_0 \oplus \mathcal{F}_1, 
\ee
where the grading zero and one is assigned respectively. Now we can define $Q$  (and in general also all higher $L_n$) to be non-zero only for inputs in $\mathcal{F}_0$ and producing an output in $\mathcal{F}_1$. Similarly, $\omega$ should be non-vanishing only if one input is in  $\mathcal{F}_0$ and one input in  $\mathcal{F}_1$. As a result, the relations \ref{L inf def}  are all satisfied identically purely for degree reasons. The fact that they are not restrictive at all is due to the lack of any non-trivial gauge symmetry in this simple example. 
\\
For later purposes let us define a basis $e_i$, $e^j$  of  $\mathcal{F}_0$ and  $\mathcal{F}_1$, respectively, such that objects like $Q_\p$ and $\sigma$ can be expanded in components. Moreover, we can diagonalize $\omega$ such that 
\be
\omega(e_i,e^j)=-\omega(e^j,e_i)=\delta^j_i.
\ee
This concludes our short review of $L_\infty$-algebras.
\subsection{The covariant phase space formalism\label{CPSF}}
The covariant phase space formalism is an elegant and powerful method to determine the phase space and its symplectic structure for generally covariant field theories \cite{Iyer:1994ys,Wald:1990mme}. Consider such a field theory and denote its Lagrangian top form by $\tb{L}$\footnote{Differential forms in spacetime will be denoted by bold-faced letters.}. We will also allow for a Gibbons-Hawking-like boundary term written as $\bds d\tb{l}$ for some $(d-1)$-form $\bds{l}$ such that the total action of the theory becomes 
\be
S=\int \tb{L}+\tb{dl}
\ee
Varying the Lagrangian, or equivalently, taking the exterior derivative in field space (see Appendix (\ref{app1}) for a review of the variational bi-complex) yields the decomposition 
\be \label{LV}
\delta \tb{L}= \tb{E}_i \delta \p^i - \tb{d}\bold{\th}.
\ee
Here, $\tb{E}_i$ is the equation-of-motion (EoM) $(d,0)$-form \footnote{In our convention \tb{d} and $\dt$ anticommute.} and $\boldsymbol{\Theta}$ is a $(d-1,1)$-form known as the \textsl{pre-symplectic potential}\footnote{Here, a \((p,q)\)-form denotes a \(p\)-form in spacetime and a \(q\)-form in configuration space.}. The multi-index $i$  is running over all dynamical fields present in the theory and includes all index contractions. The second term in the above equation arises from the integration by parts needed to remove derivatives acting on $\delta \phi$. Notice that $\mathbf{E}_i$ is uniquely determined from \eqref{LV}, whereas $\boldsymbol{\Theta}$ is determined only up to an addition of a closed $(d-1,1)$-form. In section (\ref{BBSF1}), we will see how this ambiguity can be fixed using  Anderson's homotopy operator \cite{Anderson1992,Barnich:2001jy}.
To have a well-defined variational principle, we need all the terms which localize on the spatial boundary to vanish after taking the integral. Basically, we demand 
\be
\dt S=\int_{\Sigma_+} \Psi_+ -\int_{\Sigma_-}\Psi_-
\ee
where $\Psi_\pm$ is some quantity localized on the temporal  boundary $\Sigma_\pm$, respectively\footnote{The boundary conditions at spatial infinity are part of the definition of the theory, whereas the boundary conditions at future and past  boundaries fix the state in the theory. We allow variations that change the state and therefore do not impose boundary conditions at future or past boundaries.}. Hence, if $\Gamma$ denotes the spatial boundary, we must have 
\\
\begin{figure}
    \centering
    \includegraphics[width=0.5\linewidth]{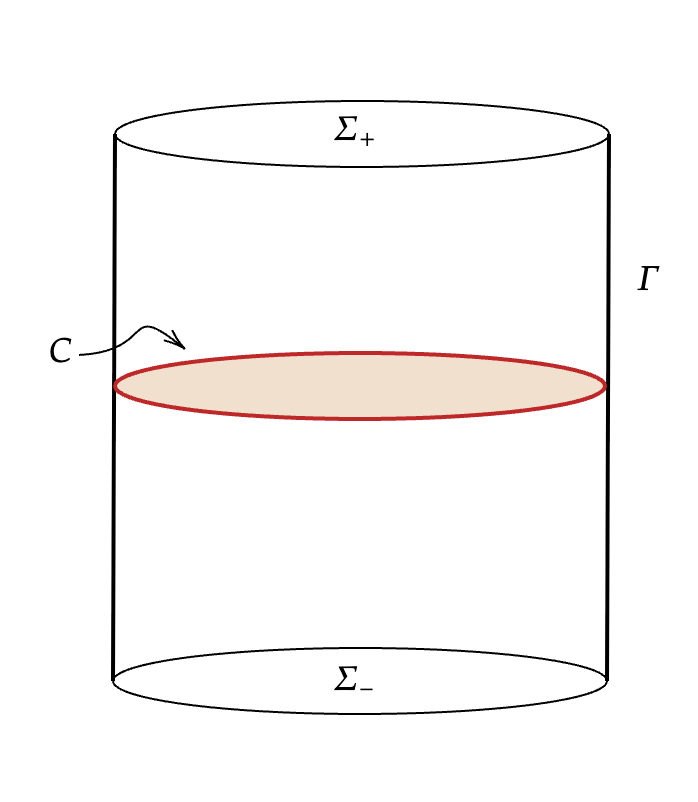}
    \caption{In this figure, \(\Sigma_+\) and \(\Sigma_-\) denote the future and past tempor al boundaries, respectively. \(\Gamma\) represents the spatial boundary. The Cauchy surface is shown in beige color, while its boundary is depicted in red, where the corner term \(C\) contributes. }
    \label{fig:corner1}
\end{figure}
\\
\be\label{boundary cond}
(\bold{\th}+\delta \tb{l})|_\Gamma =\tb{dC}
\ee
for some $(d-2,1)$-form $\tb{C}$, see figure (\ref{fig:corner1}). Typically this requires boundary conditions on the fields $\p^i$ together with a corresponding choice of $\tb{l}$. For a more detailed discussion see \cite{Harlow:2019yfa}.
Now with the help of the pre-symplectic potential $\bds{\th}$ we can define the \textsl{pre-symplectic current} as 
\be \label{sym current}
\bds{\omega}=\delta\bold{\th}-\delta\tb{dC} 
\ee
By definition, $\bds{\omega}$ is $\delta$-closed and we can show that it is also $\tb{d}$-closed on-shell:
\be
\tb{d}\bds{\omega}=\tb{d}\delta\bold{\th}-\tb{d}\delta\tb{dC}=-\delta\tb{d}\bold{\th}=-\delta \tb{E}_i \wedge \delta\p^i=0
\ee
Moreover, from \ref{boundary cond} we deduce that $\bds{\omega}$ vanishes on the spatial boundary $\Gamma$.
Let $\sg$ be some codimension-1 Cauchy surface. The \textsl{pre-symplectic form} is then defined as
\be \label{Sf}
\Tilde{\Omega}_{\sg}\eqv\int_{\sg} \bds{\omega}.
\ee
We can show that $\Tilde{\Omega}_{\sg}$ is actually independent of $\Sigma$: Consider the spacetime region $X$ bounded by two Cauchy surfaces $\Sigma_1$ and $\Sigma_2$ and possibly some spatial boundary $\Gamma'$. Integrating $\tb{d}\bds{\omega}=0$ over $X$ yields
\be\label{symplflux} 
0=\int_X  \tb{d}\bds{\omega}=\int_{\Sigma_1} \bds{\omega}-\int_{\Sigma_2} \bds{\omega}+\int_{\Gamma'} \bds{\omega} 
\Rightarrow
\Tilde{{\Omega}}_{\sg_1}=\Tilde{{\Omega}}_{\sg_2}
\ee
using the above results. 
\\
The pre-symplectic form is, in general, degenerate in the presence of gauge symmetries, which is why it is referred to as ``pre''. Namely, it is annihilated by any vector field $\Lambda$ that induces a gauge transformation,
\be
i_\Lambda {\Omega} =0.
\ee
Using Cartan's magic formula we can also see that ${\Omega}$ is constant under the flow of $\Lambda$ and therefore gauge independent:
\be\label{gauge ind}
\mathcal{L}_\Lambda {\Omega}=\delta i_\Lambda {\Omega}+i_\Lambda \delta {\Omega}=0
\ee 
To obtain a non-degenerate symplectic form we must mod out by those  zero-modes of $\Tilde{\Omega}$. We therefore fix a gauge slice, i. e. a submanifold of the pre-phase space $\Tilde{\mathcal{F}}$ that intersects each gauge slice exactly once. Then the symplectic form ${\Omega}$ is just given by the pull-back of 
$\Tilde{{\Omega}}$ to the gauge slice. The resulting form ${\Omega}$
is now by construction non-degenerate and also independent of the choice of gauge slice because of \ref{gauge ind}.
\subsection{The BEF symplectic form}\label{BEF}
\begin{figure}[h] 
  \centering
  \includegraphics[width=0.85\textwidth]{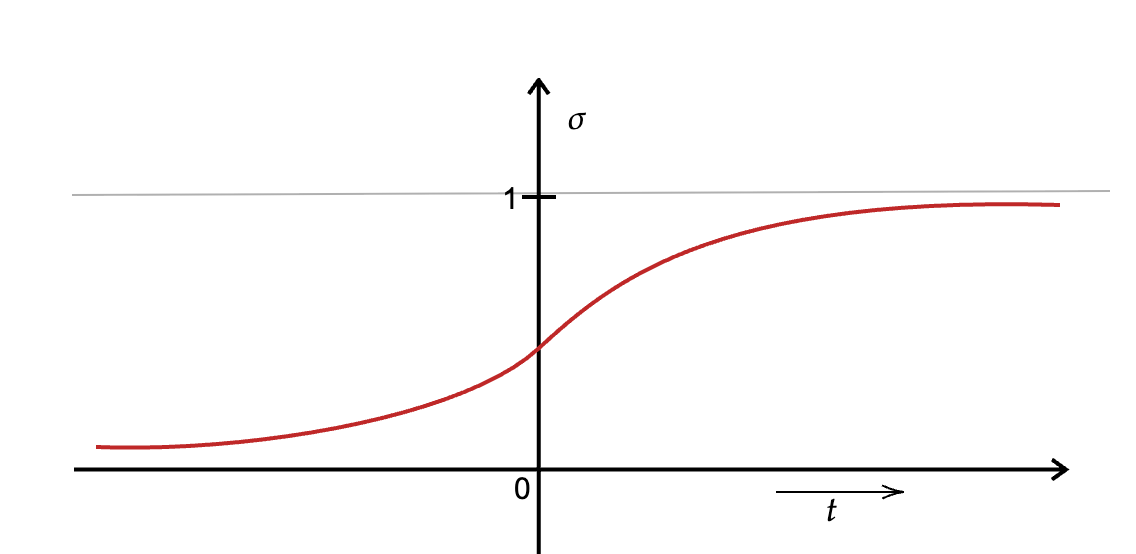}
  \caption{This figure depicts the generic behavior of the sigmoid function in an appropriately chosen time coordinate \(t\).
 }
 \label{fig:sigmoid}
\end{figure}

Bernardes, Erler and F\i rat in \cite{Bernardes:2025uzg} proposed the following expression for the symplectic form on phase space for any field theory formulated in $L_\infty$-language:
\be\label{BEF sym}
\Omega_{BEF}=\frac{1}{2}\omega(\dt \p,[Q_ \p,\sigma]\dt\p).
\ee
\(\sigma\) is called the \textsl{sigmoid operator} and satisfies the boundary conditions
\begin{align}\label{sigm bdy}
\lim_{t\rightarrow -\infty}\sigma&=0 && \lim_{t\rightarrow \infty}\sigma=1 
\end{align}
as well as 
\be
\omega(\p_1,\sigma \p_2)=\omega(\sigma\p_1, \p_2).
\ee
Apart from that it is basically unconstrained, but in our examples it will just be given as a function of spacetime acting by multiplication. We do not restrict it to be constant in space though but just a general, covariant function, see Figure~(\ref{fig:sigmoid}) . Since the commutator in (\ref{BEF sym}) vanishes at early and late times, the domain of $\Omega_{BEF}$ is a kind of spread, diffuse Cauchy splice, namely the region where $\sigma$ is changing. In the simplest case,  $\sigma$ is just a step function $H(t-t_0$) and at least in local theories, $\Omega_{BEF}$ is localized to a codimension-1 Cauchy surface.\footnote{In non-local theories like string field theory this does not need to be true, see \cite{Bernardes:2025uzg, Bernardes:2025zzu}.} 
\\
Following \cite{Bernardes:2025uzg}, we note that $\Omega_{BEF}$ is naively zero because of \ref{Q_phi kills delta phi} and cyclicity of $Q_\p$. However, cyclicity holds only up to total derivatives and requires a careful treatment of the temporal boundary contributions, which is illustrated by the use of $\tau$-regularization. Related to that one should notice that the two terms in the commutator naively give the same contribution, as long as boundary terms are ignored:
\be
\omega(\dt \p,Q_ \p\sigma\dt\p)=\omega(Q_\p\dt \p,\sigma\dt\p)=\omega(\sigma Q_\p\dt \p,\dt\p)=-\omega(\dt \p,\sigma Q_ \p\dt\p)
\ee
Though, for the above reason it makes sense to write the expression as a commutator and refrain from splitting it into parts within the bilinear form. In section \ref{cons} we will give a more precise argument for that.
\\
 For being a valid symplectic form, it must be $\delta$-closed, gauge invariant and independent of the choice of $\sigma$. All those properties were proven in \cite{Bernardes:2025uzg} and will be revisited in section \ref{Sb11} in presence of a spatial boundary. Moreover, the consistency of $\Omega_{BEF}$ was illustrated in various examples, including non-local ones \cite{Bernardes:2025uzg}.
 \\
 \section{Relation of $\Omega_{BEF}$ to the covariant phase space formalism}\label{main proof}
We will now show the consistency of ${\Omega}$ from the covariant phase space formalism and $\Omega_{BEF}$  in the absence of any spatial boundary. This proof is one of the main results so we will present it in detail. 
\\
First, from \ref{sym current}  and \ref{Sf} one can see that the term containing $C$ localizes on $\partial\Sigma$, i. e. on the "corner" where the Cauchy surface intersects the spatial boundary. Those corner terms will become relevant in section \ref{Sb11}, where we will consider a spatial boundary, but do not contribute for now. The idea of the proof is roughly speaking the following: We show that the commutator of $Q_\p$ 
with $\sigma$ basically produces the same terms as partial integration in the covariant formalism. Here we assume that cyclicity holds only up to total derivatives. Let us consider the following modified action in $L_\infty$- form:
\be
S_\sigma= -\frac{1}{2}\omega(\sigma \phi,Q\phi)-\sum_{n=2}^\infty \frac{1}{(n+1)!}\omega(\sigma \phi,L_n(\phi,...,\phi))
\ee
The insertion of $\sigma$ has the meaning of "turning on" the fields at some finite time and therefore creating a fuzzy, temporal boundary. Hence, from \ref{LV} we expect that the $\bds d\bds\Theta$-term will localize in the region where $\sigma$  is changing. In this way, the construction remains fully compatible with our understanding of covariant phase space methods in local field theories. To obtain the corresponding Lagrangian, observe that for any covariant field theory, $\omega$ contains the integral over spacetime, see \ref{free scalar ex}. Let us therefore define the \textsl{reduced bilinear form} $\omega'$ via 
\be
\omega (\p_1,\p_2)=\int_\mathcal{M}\bds{ \omega'} (\p_1,\p_2).
\ee
In this way, $\bds{ \omega'}$ becomes a top form in spacetime and we can use all the formalism of section \ref{CPSF}.
Now, let us vary the Lagrangian
\begin{multline}
\delta \bds{ L_\sigma}=  \frac{1}{2}\bds{ \omega'}(\sigma \dt\phi,Q\phi)-\frac{1}{2}\bds{ \omega'}(\sigma \phi,Q\dt\phi)+\sum_{n=2}^\infty\frac{1}{(n+1)!}\bds{ \omega'}(\sigma \dt\phi, L_n(\phi,...,\phi))\\+\sum_{n=2}^\infty\frac{1}{(n+1)!}\bds{ \omega'}(\sigma \phi,\dt L_n(\phi,...,\phi))
\end{multline}
where we have used \(\dt\omega(A,B)=-\omega(\dt A,B)-(-1)^{|A|}\omega(A,\dt B)\). Since \(\dt L_n(\phi,..\phi)=-nL_n(\phi,..,\phi,\dt\phi)\) we get
\begin{multline}
\delta\bds{ L_\sigma}=  \frac{1}{2}\bds{ \omega'}(\sigma \dt\phi,Q\phi)+\sum_{n=2}^\infty\frac{1}{(n+1)!}\bds{ \omega'}(\sigma \dt\phi, L_n(\phi,...,\phi))-\frac{1}{2}\bds{ \omega'}(\sigma \phi,Q\dt\phi)\\-\sum_{n=2}^\infty\frac{n}{(n+1)!}\bds{ \omega'}(\sigma \phi,L_n(\phi,...,\dt\phi)).
\end{multline}
Using cyclicity of the \(L_n\) up to total derivatives, which we denote as \(\bds{d\Theta_\sigma}\), we get
\begin{multline}
\delta \bds{ L_\sigma}=  \frac{1}{2}\bds{ \omega'}(\sigma \dt\phi,Q\phi)+\sum_{n=2}^\infty\frac{1}{(n+1)!}\bds{ \omega'}(\sigma \dt\phi, L_n(\phi,...,\phi))+\frac{1}{2}\bds{ \omega'}(Q(\sigma \phi),\dt\phi)\\+\sum_{n=2}^\infty\frac{n}{(n+1)!}\bds{ \omega'}(L_n(\sigma \phi,...,\phi),\dt\phi)-\bds{d\Theta_\sigma}.
\end{multline}
 Now, using \ref{antisym omega} we can write it as
\begin{multline}
\delta \bds{ L_\sigma}=  \frac{1}{2}\bds{ \omega'}(\sigma \dt\phi,Q\phi)+\sum_{n=2}^\infty\frac{1}{(n+1)!}\bds{ \omega'}(\sigma \dt\phi, L_n(\phi,...,\phi))+\frac{1}{2}\bds{ \omega'}(\dt\phi,Q(\sigma \phi))\\+\sum_{n=2}^\infty\frac{n}{(n+1)!}\bds{ \omega'}(\dt\phi,L_n(\sigma \phi,...,\phi))-\bds{d\Theta_\sigma}.
\end{multline}
Since the $L_n$ are totally graded symmetric, the expression becomes
\begin{multline}
\delta \bds{ L_\sigma}=  \bds{ \omega'}(\sigma \dt\phi,Q\phi)+\sum_{n=2}^\infty\frac{1}{n!}\bds{ \omega'}(\sigma \dt\phi, L_n(\phi,...,\phi))+\frac{1}{2}\bds{ \omega'}(\dt\phi,[Q,\sigma] \phi))\\+\sum_{n=2}^\infty\frac{n}{(n+1)!}\bds{ \omega'}(\dt\phi,[L_n(\phi,...,\phi,\cdot ),\sigma]\phi)-\bds{d\Theta_\sigma}.
\end{multline}
Now, in the first two terms we can recognize the equations of motion \(E[\phi]=Q\phi+\sum_{n=2}\frac{1}{n!}L_n(\phi,...,\phi)\):
\begin{multline}
\delta \bds{ L_\sigma}=  \bds{ \omega'}(\sigma \dt\phi,E[\phi])+\frac{1}{2}\bds{ \omega'}(\dt\phi,[Q,\sigma] \phi)+\sum_{n=2}^\infty\frac{n}{(n+1)!}\bds{ \omega'}(\dt\phi,[L_n(\phi,...,\phi,\cdot ),\sigma]\phi)-\bds{d\Theta_\sigma}.
\end{multline}
One should note at this point that the cyclicity manipulations we used are exactly the same as needed to derive the equations of motion for the ordinary action \ref{L inf action}. Therefore, the total derivative term $\bds{d\Theta_\sigma}$ indeed corresponds to the $\bds{d\Theta}$ term in \ref{LV}. Let us now take another variation of the above expression:
\begin{multline}
0=  \bds{ \omega'}(\sigma \dt\phi,\dt E[\phi])-\frac{1}{2}\bds{ \omega'}(\dt\phi,[Q,\sigma] \dt\phi)+\sum_{n=2}^\infty\frac{n}{(n+1)!}\bds{ \omega'}(\dt\phi,\dt\{[L_n(\phi,...,\phi,\cdot ),\sigma]\phi\})-\dt \bds{d\Theta_\sigma}
\end{multline}
therefore,
\begin{multline}
-\dt \bds{d\Theta_\sigma}= - \bds{ \omega'}(\sigma \dt\phi,\dt E[\phi])+\frac{1}{2}\bds{ \omega'}(\dt\phi,[Q,\sigma] \dt\phi)-\sum_{n=2}^\infty\frac{n}{(n+1)!}\bds{ \omega'}(\dt\phi,\dt\{[L_n(\phi,...,\phi,\cdot ),\sigma]\phi\}).
\end{multline}
At this time we can already observe something interesting: Assuming the equations of motion to hold, both terms on the r. h. s. contain a commutator with $\sigma$, this means after taking the spacetime integral, they will localize in the region where $\sigma$ is changing, which was expected. 
\\
Let us now manipulate the last term: Since we already know that the term localizes on the generalized Cauchy surface, any total derivative we pick up would contribute only at the spatial boundary, i. e. on the (generalized) corners. We assume that no spatial boundary is present at the moment, we can discard those terms and freely use cyclicity\footnote{The effects of a non-trivial spatial boundary will be discussed in \ref{Sb11}: To use cyclicity, it is necessary to introduce a covariantized version of $\tau$-regulation.}:  
\begin{align*}
&\bds{ \omega'}(\dt\phi,\dt\{[L_n(\phi,...,\phi,\cdot ),\sigma]\phi\})=\bds{ \omega'}(\dt\phi,\dt\{L_n(\phi,...,\phi,\sigma \phi)-\sigma L_n(\phi,...,\phi)\}\\
    =&-\bds{ \omega'}(\dt\phi,(n-1)L_n(\phi,...,\dt\phi,\sigma \phi)) - \bds{ \omega'}(\dt\phi,L_n(\phi,...,\phi,\sigma \dt\phi)) +n\bds{ \omega'}(\dt\phi,\sigma L_n(\phi,...,\phi,\dt\phi)\})\\
    =&-(n-1)\bds{ \omega'}(L_n(\dt\phi,\phi,..,\dt\phi),\sigma \phi)-\frac{(n+1)}{2}\bds{ \omega'}( \dt\phi,[L_n(\phi,\phi,..,\phi,\cdot ),\sigma]\dt \phi)\\& \ \ \ \ \ +\frac{n-1}{2}\Big(\bds{ \omega'}(L_n(\dt\phi,\phi,..,\phi),\sigma \dt\phi)+\bds{ \omega'}(\dt\phi,\sigma L_n(\phi,...,\phi,\dt\phi))\Big)\\
    =&-\frac{(n+1)}{2}\bds{ \omega'}( \dt\phi,[L_n(\phi,\phi,..,\phi,\cdot ),\sigma]\dt \phi)
\end{align*}
We used the fact that \(L_n(\dt\phi,\phi,..,\phi,\dt\phi)=0\) and also noticed that the two terms in the fourth line cancel each other. Therefore, by inserting the above equation and also assuming to be on-shell we get 

\begin{multline}
-\dt \bds{d\Theta_\sigma}= \frac{1}{2}\bds{\omega'}(\dt\phi,[Q,\sigma] \dt\phi)+\frac{1}{2}\sum_{n=2}^\infty\frac{1}{(n-1)!}\bds{\omega'}(\dt\phi,[L_n(\phi,...,\phi,\cdot ),\sigma]\dt \phi)
\end{multline}
At this point we can recognize the operator $Q_\p$ defined in \ref{Q_phi def} and write 
\be
-\dt \bds{d \Theta_\sigma}= \frac{1}{2}\bds{ \omega'}(\dt\phi,[Q_\p,\sigma] \dt\phi).
\ee
Reintroducing the spacetime integral basically gives the desired result:
\be\label{Omega equivalence}
-\int_\mathcal{M} \dt \bds d\bds{\Theta_\sigma}= \frac{1}{2}\omega(\dt\phi,[Q_\p,\sigma] \dt\phi) =\Omega_{BEF}.
\ee
To make contact with \ref{Sf}, let us assume we have a Cauchy surface $\Sigma$ defined as the zero locus of some characteristic function $f$:
\be
\Sigma = \{x^\mu|f(x^\mu)=0\}
\ee
Then \ref{Sf} can be alternatively written as
\be
\Tilde{\Omega}_{\sg}\eqv-\int_{\mathcal{M}}\dt \bds{d}(\bds{\Theta} H(f(x))) \ ,
\ee
(again, ignoring the corner terms). After comparing with \ref{Omega equivalence} we see that in $\Omega_{BEF}$ the Heaviside function just got replaced by a generalized, covariant localization induced by the function $\sigma$. In this sense, $\Omega_{BEF}$ computes the "weighted average" of symplectic forms $\Tilde{\Omega}_{\sg}$  over a family of Cauchy surfaces $\Sigma$.

\section{Adding a spatial boundary}\label{Sb11}
In this section, we establish the consistency of the BEF symplectic form in the presence of spatial boundaries. We then briefly define the Barnich-Brandt symplectic form and derive its relation to the BEF symplectic form.
\subsection{Consistency of $\Omega_{BEF}$}\label{cons} 
\begin{figure}[h!]
\centering
\begin{subfigure}{0.6\textwidth}
    \centering
    \includegraphics[width=\linewidth]{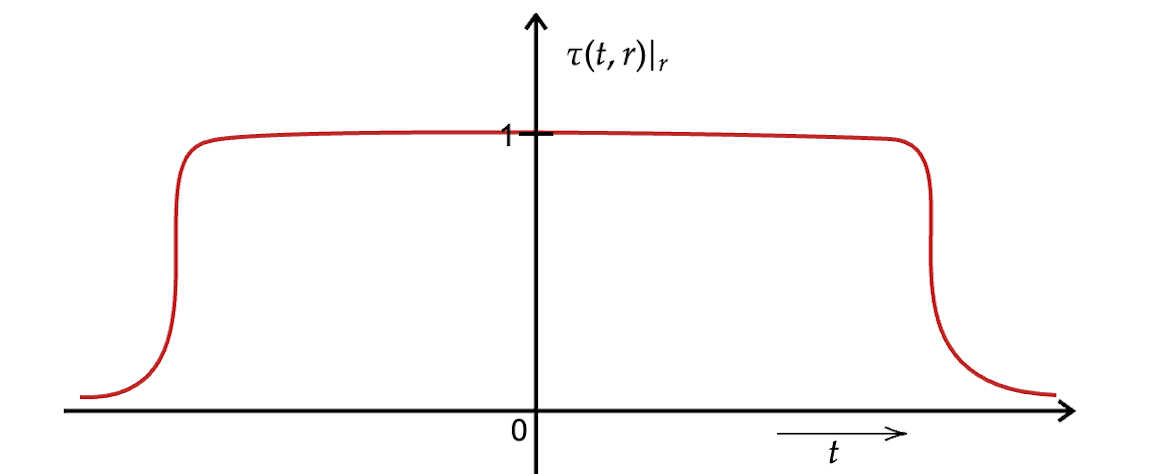}
    \caption{This figure depicts the temporal behaviour of the \(\tau\)-regulator at fixed r.}
\end{subfigure}
\hfill
\begin{subfigure}{0.35\textwidth}
    \centering
    \includegraphics[width=\linewidth]{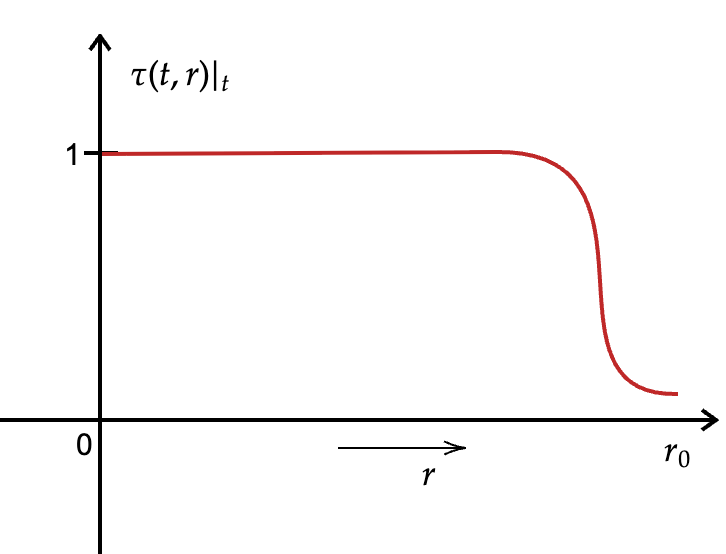}
    \caption{This figure depicts the spatial behaviour of the \(\tau\)-regulator at fixed time.}
\end{subfigure}
\caption{The above figures depict the behaviour of the \(\tau\)-regulator. The spatial boundary is located at \(r_0\) in suitables coordinates.}
\label{fig:tau}
\end{figure}

When introducing a non-trivial spatial boundary, first of all we have to review and check the necessary properties of $\Omega_{BEF}$  in this new context. This means, we have to show that it is still $\dt$-closed, gauge invariant and independent of $\sigma$. An important tool for that, introduced in \cite{Bernardes:2025uzg}, is the $\tau$-\textsl{regulator}. Originally, it is defined similar to the sigmoid, but cutting off contributions from both very early and very late times. Hence it fulfills the "boundary conditions" $\tau \rightarrow0$  for $t\rightarrow\pm\infty$ and $\tau=1$ for finite $t$. This is not a mathematically rigorous definition, but in practice it is useful to regularize the (otherwise infinite) integral over time. Now, having a spatial boundary as well, it makes sense to "covariantize" $\tau$, such that it also obeys  $\tau \rightarrow0$  for $r$ approaching the boundary value $r_0$ and $\tau=1$ otherwise, for some radial coordinate $r$.\footnote{We assume that spacetime has the trivial topology $\mathbb{R}\times\Sigma$.}. The behaviour of the \(\tau\)- regulator is shown in figure (\ref{fig:tau}). Similar to $\sigma$, we also assume $\tau$ to be commuting and fulfill 
\be
\omega(\p_1,\tau \p_2)=\omega(\tau\p_1, \p_2).
\ee

Let us now prove the desired properties using the covariant $\tau$-regulator:
\subsubsection{Closedness}
First of all, let us define a prescription for applying $\tau$. The symplectic form is defined around some background solution $\p$  that should not be affected by any regularization. Instead we will define a new variation $\dt_\tau\p=:\tau\dt\p$.  This will have the effect, despite of regularizing the time integral, to create a spatial boundary at $r=r_0$ for the perturbations around $\p$. We get 
\be
\dt_\tau \Omega_{BEF}=\frac{1}{2}\omega(\tau\dt\phi,[\dt_\tau Q_{\p},\sigma]\tau \dt\phi)=-\frac{1}{2}\omega(\tau\dt\phi,[L_2^{\p}(\tau\dt\p,\cdot),\sigma]\tau \dt\phi)
\ee
where we have introduced the operator
\be
L_2^\p(\p_1,\p_2)=\sum_{n=2}^\infty \frac{1}{(n-2)!}L_n(\p_1,\p_2,\p,...,\p)
\ee
in a way similar to $Q_\p$. $L_2^\p$  is cyclic, as can be easily seen, so we can write 
\be
\dt_\tau \Omega_{BEF}=\frac{1}{2}\omega(L_2^{\p}(\tau\dt\phi,\tau\dt\p),\sigma\tau \dt\phi)+\frac{1}{2}\omega(\tau\dt\phi,\sigma L_2^{\p}(\tau\dt\p,\tau \dt\phi))=0
\ee
because $L_2^\p$ is graded symmetric.

\subsubsection{Gauge invariance}
We have to show that the action of $\Omega_{BEF}$ on vector fields $v_\Lambda$ 
that generate a gauge transformation vanishes:
\be
i_{v_\Lambda}\Omega_{BEF}=0
\ee
From \ref{Q_phi def}  and the related discussion as well as our prescription for applying $\tau$, we know that such vector fields take the form $v_\Lambda =\tau Q_\p \Lambda$, hence we get  
\begin{align*}
   i_{v_\Lambda}\Omega_{BEF}&=-\omega(\tau Q_\p \Lambda,[Q_ \p,\sigma]\tau\dt\p)=-\omega(\tau\Lambda,Q_\p ([Q_ \p,\sigma]\tau\dt\p))+\omega([Q_\p,\tau] \Lambda,[Q_ \p,\sigma]\tau\dt\p)\\
  & =-\omega(\tau\Lambda, [Q_ \p^2,\sigma]\tau\dt\p)+\omega(\tau\Lambda, [Q_ \p,\sigma]Q_\p\tau\dt\p))+\omega([Q_\p,\tau] \Lambda,[Q_ \p,\sigma]\tau\dt\p)\\&=\omega([Q_\p,\tau] \Lambda,[Q_ \p,\sigma]\tau\dt\p)+\omega(\tau\Lambda, [Q_ \p,\sigma][Q_\p,\tau]\dt\p))
   \end{align*}
since $Q_\p$ is nilpotent on-shell and \ref{Q_phi kills delta phi}  holds. Hence, we have produced a non-vanishing term that localizes on the boundary where $\tau$ is changing. The easiest way to kill this term, and also the way we will adopt here, is to restrict $\Lambda$ to have compact support in the bulk. This means, $\Lambda$  and all its derivatives will vanish in the region where $\tau$  is changing and the above term vanishes, at least for local theories. This is a common requirement that excludes topological non-trivial, "large" gauge transformations such that the fields only transform locally in the bulk.\footnote{The analysis of large gauge transformations will be left for future work.} The fate of gauge symmetry in the presence of a boundary is actually a well-known issue. An alternative strategy to solve it was pursued for instance in \cite{Maccaferri:2025orz, Maccaferri:2025onc} and requires additional fields living on the boundary. The implementation is dependent on the theory in question though and would lead beyond the scope of this paper. For non-local theories, if any issue arises, we can always define small gauge transformations those for which $i_{v_\Lambda}\Omega_{BEF}=0$. 
Full gauge invariance now follows directly from closedness and Cartan's magic formula:
\be
\mathcal{L}_{v_\Lambda} \Omega_{BEF}=\delta i_{v_\Lambda}\Omega_{BEF}+i_{v_\Lambda} \delta \Omega_{BEF}=0
\ee
\subsubsection{Independence of $\sigma$} 
Finally we have to show that $ \Omega_{BEF}$ is independent of the sigmoid or, equivalently, that symplectic flux is conserved. Given two different sigmoids we can write 
\begin{align}
    \Omega_{BEF}-\Omega_{BEF}'=\frac{1}{2}\omega(\tau\dt\p,[Q_{\p},(\sigma-\sigma')]\tau\dt\p).
\end{align}
Using the cyclicity of $Q_{\p}$  this expression becomes 
\begin{align*}
  \frac{1}{2}\omega([Q_{\p},\tau]\dt\p,(\sigma-\sigma')\tau\dt\p)-
  \frac{1}{2}\omega(\tau\dt\p,(\sigma-\sigma')[Q_{\p},\tau]\dt\p)= \omega([Q_{\p},\tau]\dt\p,(\sigma-\sigma')\tau\dt\p)
\end{align*}
where we have already used $Q_\p\dt\p=0$. For general covariant theories, $[Q_{\p},\tau]$ will have precisely the same structure as $[Q_{\p},\sigma]$, just localized to the spatial boundary instead of the space-like Cauchy slice. This means, the above expression is just $\Omega_{BEF}$  localized on the spatial boundary region $\Gamma'$ between $\pd\Sigma_1$ and $\pd\Sigma_2$. It equals the symplectic flux contribution in \ref{symplflux}, which is typically set to zero by the boundary conditions.\footnote{In a more general setting, the flux might be non-zero and an important quantity to compute, for instance, if the boundary is some sort of interface rather than a physical boundary. Also for some theories, especially non-local theories like string field theory, it is not known how to apply consistent boundary conditions.}  In the case where  $\pd\Sigma_1=\pd\Sigma_2$, we trivially get $  \Omega_{BEF}=\Omega_{BEF}'$.  
\\
Also note that the proof of section \ref{main proof} will go through after specifying boundary conditions and adding an appropriate boundary Lagrangian, following section \ref{CPSF}.
\subsection{The Barnich-Brandt symplectic form}\label{BBSF1}
Let us recall equation \ref{LV}  and the fact that the pre-symplectic potential $ \bds\Theta$ was determined from the Lagrangian only up to some $(d-2,1)$-form,
\be
\bds\Theta \sim \bds\Theta+\bds d \bds C.
\ee
After integrating the pre-symplectic current over the Cauchy surface $\Sigma$,  the $\bds C$-term will localize on the boundary of $\Sigma$, i. e. on the corners where $\Sigma$  intersects the spatial boundary. Therefore, in the previous analysis without considering any spatial boundary, we did not have to care about this ambiguity. In this case, $\bds\omega=\dt\bds\Theta$  is known as the \textsl{Iyer-Wald pre-symplectic current}. Now, the $\bds C$-term becomes relevant and Barnich and Brandt \cite{Barnich:2001jy, Compere:2018aar} have found an elegant way to determine $\bds \Theta$ uniquely using the Anderson homotopy operator \ref{homoptopy operator}. As we will see the resulting symplectic current will differ from the Iyer-Wald current by some canonical corner term.\footnote{If this term contributes or not, depends on the boundary conditions.}  For instance, in the case of General Relativity, it yields the only possible, covariant corner term \cite{Compere:2018aar}. Also, it plays an important role in the study of asymptotic algebras, see \cite{Barnich:2001jy}. Let us apply equation (\ref{homtopy alg1}) on the Lagrangian $\bds L$: 
\be\label{hom V}
\bds{\dt}\tb{L}=\dt\p\frac{\dt \tb{L}}{\dt\p}-\tb{d}I^n_{\dt\p}\tb{L}
\ee
We know that the Euler operator $\frac{\dt }{\dt\p}$ just produces the equations of motion, hence comparing with \ref{LV}  we see that the above equation fixes \(\bds{\th}=I^d_{\dt\p}\tb{L}\). This gives us an algorithm to determine the pre-symplectic potential from the Lagrangian without any ambiguity.  The idea behind is that integration by parts can be made a  unique, well-defined operation by symmetrizing over all inequivalent ways. Further,
\be
    I^d_{\dt\p}\bds{\dt}\tb{L}= I^d_{\dt\p}(\dt\p\frac{\dt \tb{L}}{\dt\p})- I^d_{\dt\p}\tb{d}I^d_{\dt\p}\tb{L}
\ee
and using (\ref{homtopy alg2}) we get
\be
    I^d_{\dt\p}\bds{\dt}\tb{L}= I^d_{\dt\p}(\dt\p\frac{\dt \tb{L}}{\dt\p})- \bds{\dt}I^d_{\dt\p}\tb{L}-\tb{d}I^{d-1}_{\dt\p}I^{d}_{\dt\p}\tb{L}.
\ee
The fact that \([\bds{\dt},I^n_{\dt\p}]=0\)  helps us to simplify that and get 
\be
\bds{\dt}I^d_{\dt\p}\tb{L}= \frac{1}{2}I^d_{\dt\p}(\dt\p\frac{\dt \tb{L}}{\dt\p})-\frac{1}{2}\tb{d}I^{d-1}_{\dt\p}I^{d}_{\dt\p}\tb{L}.
\ee
The left hand side is just the Iyer-Wald pre-symplectic current $\bds\omega^{IW}$ and the first term of the right hand side will be defined as \textsl{Barnich-Brandt symplectic current} $\bds\omega^{BB}$. They differ by a total derivative term  
\be\label{BB IW relation}
\bds{\omega}^{BB}(\p;\dt\p,\dt\p)=\bds{\omega}^{IW}(\p;\dt\p,\dt\p)-\tb{dB}(\p;\dt\p,\dt\p)
\ee
where
\begin{align}\label{f1}
 \tb{B}(\p;\dt\p,\dt\p)=-\frac{1}{2}I^{d-1}_{\dt\p}I^{d}_{\dt\p}\tb{L}
\end{align}
This $(d-2,2)$-form is precisely what will localize on the corners after integrating over $\Sigma$. Note that $\boldsymbol{\omega}^{BB}$ is defined directly in terms of the equations of motion, hence it is invariant under all standard ambiguities in the Lagrangian and in the pre-symplectic potential. 

\subsection{$\Omega_{BEF}$ and $\bds{ \omega^{BB}}$ in second order theories }
We now want to examine the precise relation between the Barnich-Brandt symplectic form and $\Omega_{BEF}$ and start with theories of at most second order equations of motion. This includes almost all fundamental theories and will illustrate the important steps in a transparent way.\\
We assume that $\frac{\delta \mathbf{L}}{\delta \phi}$ contains at most second derivatives of $\phi$, and thus the action of $I^{n}_{\delta \phi}$ truncates at second order in derivatives of $\phi$. Therefore, we can apply the formula given in equation (\ref{ex homtopy1}) and forget about all the terms in the ellipsis:
\begin{align*}
-I^n_{\delta \phi} \Big( \delta \phi^i \frac{\delta \mathbf{L}}{\delta \phi^i} \Big)
&= \delta \phi^j \wedge \delta \phi^i 
   \frac{\partial}{\partial \phi^j_{,a}} \Big( \frac{\delta \mathbf{L}_a}{\delta \phi^i} \Big)
- \delta \phi^j \wedge \delta \phi^i \, \partial_b 
   \frac{\partial}{\partial \phi^j_{,ab}} \Big( \frac{\delta \mathbf{L}_a}{\delta \phi^i} \Big) \\
&\quad - \delta \phi^j \wedge \partial_b \delta \phi^i 
   \frac{\partial}{\partial \phi^j_{,ab}} \Big( \frac{\delta \mathbf{L}_a}{\delta \phi^i} \Big)
+ \delta \phi^j_{,b} \wedge \delta \phi^i 
   \frac{\partial}{\partial \phi^j_{,ab}} \Big( \frac{\delta \mathbf{L}_a}{\delta \phi^i} \Big)\\
   &= \delta \phi^j \wedge \delta \phi^i 
   \frac{\partial}{\partial \phi^j_{,a}} \Big( \frac{\delta \mathbf{L}_a}{\delta \phi^i} \Big)
- \delta \phi^j \wedge \delta \phi^i \, \partial_b 
   \frac{\partial}{\partial \phi^j_{,ab}} \Big( \frac{\delta \mathbf{L}_a}{\delta \phi^i} \Big) \\
&\quad -\frac{\partial}{\partial \phi^j_{,ab}}  \Big( \frac{\delta \mathbf{L}_a}{\delta \phi^i} \Big)\Big(\delta \phi^j \wedge \partial_b \delta \phi^i 
- \partial_b \delta \phi^j \wedge \delta \phi^i \Big)
\end{align*}
where \(\tb{L}_a=i_{\partial_a}\tb{L}\)\footnote{Note that \(i_{\pd_a}\) is an anticommuting object that produces an overall minus sign.}. It follows from the action principle that \(\frac{\partial}{\partial \phi^j_{,ab}}  \Big( \frac{\delta \mathbf{L}_a}{\delta \phi^i} \Big)\) is symmetric in \(i\) and \(j\). Therefore,
\begin{align}
-I^n_{\delta \phi} \Big( \delta \phi^i \frac{\delta \mathbf{L}}{\delta \phi^i} \Big)
= \delta \phi^j \wedge \delta \phi^i 
   \frac{\partial}{\partial \phi^j_{,a}} \Big( \frac{\delta \mathbf{L}_a}{\delta \phi^i} \Big) -2\frac{\partial}{\partial \phi^j_{,ab}}  \Big( \frac{\delta \mathbf{L}_a}{\delta \phi^i} \Big)\Big(\delta \phi^j \wedge \partial_b \delta \phi^i \Big)
\end{align} 
where we can substitute $\bds{\omega^{BB}}$ on the left hand side:
\be\label{BB}
-\bds{\omega}^{BB}(\p;\dt\p,\dt\p)=\frac{\partial}{\partial \phi^j_{,ab}}  \Big( \frac{\delta \mathbf{L}_a}{\delta \phi^i} \Big)\Big(\partial_b \delta \phi^i \w \delta \phi^j \Big)+\frac{1}{2} \delta \phi^j \wedge \delta \phi^i 
   \frac{\partial}{\partial \phi^j_{,a}} \Big( \frac{\delta \mathbf{L}_a}{\delta \phi^i} \Big) .
\ee
It is useful to write above expression as
\be\label{BB2}
\bds{\omega}^{BB}(\p;\dt\p,\dt\p)=-\Big[\frac{\partial E_i }{\partial \phi^j_{,ab}}  \Big(\partial_b \delta \phi^i \w \delta \phi^j \Big)+\frac{1}{2} \delta \phi^j \wedge \delta \phi^i 
   \frac{\partial E_i}{\partial \phi^j_{,a}}  \Big]\bds{\ep}_a=\omega^a\bds{\ep}_a
\ee
where \(E_i\) is the equation of motion for the \(i^{th}\) field. \\

Now let us do a similar computation for $\Omega_{BEF}$. As already indicated in section \ref{BEF}, it is useful to choose a basis of $V$ and write $Q_\p$  and $\sigma$ in components as  
\begin{align}
Q_{\p}e_i&=Q_{ij}e^j && \sigma e_i=\sigma_i^j e_j&& \sigma e^i=\sigma^i_je^j
\end{align}
where cyclicity of \(Q_{\p}\) inside the bilinear form $\omega$  implies that $Q$ is self-adjoint. Further, if we take \(\sigma^k_j\) to act by multiplication with the scalar function \(\sigma(x)\), then \(\sigma_j^k=\dt_j^k \sigma(x)\). Now the equation (\ref{BEF sym}) reduces to
\be\label{ft BEF}
\Omega_{BEF}=-\frac{1}{2} \int \dt\p^i \w [\tb{Q}_{ij},\sigma(x)]\dt\p^j
\ee
Here, for convenience, we put the spacetime volume form into $Q_{ij}$, turning it into a top form \(\tb{Q}_{ij}\). We can deduce it directly from the action for the fluctuation,
\be\label{ft BEF 1}
S_{fluc}=-\frac{1}{2}\int \dt \p^i \tb{Q}_{ij} \dt \p^j
\ee
or from the linearized equations of motion \(-\tb{Q}_{ij}\dt\p^j=0\). \\
In any second order theory, \(\tb{Q}_{ij}\) can be expanded in derivatives as\footnote{If the fields are Lie algebra valued, the covariant derivatives will be replaced by gauge covariant derivatives. After the replacement the analysis will go through.},
\be\label{op exp}
\tb{Q}_{ij}=\tb{Q}_{ij}^{ab}\partial_a\partial_b+\tb{Q}_{ij}^{a}\partial_a+\tb{E}_{ij}.
\ee
Note that \(\tb{Q}_{ij}^{ab}\) is symmetric in \(i\) and \(j\) which follows from the Helmholtz conditions\footnote{It also follows from the fact that the  principal symbol of the equations of motion that are obtained from the Lagrangian is symmetric\cite{Ali:2025ybt}.}, see \cite{Anderson1992}, as well as in \(a\) and \(b\). \([\tb{Q}_{ij},\sigma(x)]\) will receive contributions only from terms which contain derivatives, so we calculate
\begin{align}\label{com}
    [\partial_a\partial_b,\sigma]&=\partial_a\sigma\partial_b+\partial_b\sigma\partial_a+\partial_a\partial_b\sigma && [\partial_a,\sigma]=\partial_a\sigma.
\end{align}
Also, the \(\partial_a\partial_b\sigma\) drops out of (\ref{ft BEF}) because of the symmetry in $i$ and $j$. Now, plugging (\ref{op exp}) and (\ref{com}) into (\ref{ft BEF}), we get
\be\label{final}
\Omega=-\int \partial_a\sigma\Big(\frac{1}{2}\dt\p^i \w \dt\p^j \tb{Q}_{ij}^{a} + \dt\p^i \w \partial_b\dt\p^j \tb{Q}_{ij}^{ab}\Big)=\int (\partial_a\sigma) \omega^a \bds{\ep}= \int \tb{d}\sigma \w \bds{\omega}
\ee
with
\begin{align}\label{fsym}
\omega^a&=-\Big(\frac{1}{2}\dt\p^i \w \dt\p^j \tb{Q}_{ij}^{a} + \dt\p^i \w \partial_b\dt\p^j \tb{Q}_{ij}^{ab}\Big) && \bds{\omega}=\omega^a\bds{\ep}_a
\end{align}
where \(\bds{\ep}_a=i_{\partial_a}\bds{\ep}\) is the codimension one volume form. 
\\
Now, using (\ref{vertical der}), we can write
\be
\dt\frac{\dt \tb{L}}{\dt \p^i}=\Big(\frac{\partial}{\partial\phi^j}\frac{\dt \tb{L}}{\dt \p^i}\Big) \dt\p^j+\Big(\frac{\partial}{\partial\partial_a \phi^j}\frac{\dt \tb{L}}{\dt \p^i}\Big) \partial_a\dt\p^j+\Big(\frac{\partial}{\partial\partial_a\partial_b \phi^j}\frac{\dt \tb{L}}{\dt \p^i}\Big) \partial_a\partial_b\dt\p^j
\ee
and read off
\begin{align}\label{op}
    \tb{Q}_{ij}^{ab}&=-\frac{\partial}{\partial\partial_a\partial_b \phi^j}\frac{\dt \tb{L}}{\dt \p^i} ~~~\tb{Q}_{ij}^{a}=-\frac{\partial}{\partial\partial_a \phi^j}\frac{\dt \tb{L}}{\dt \p^i}.
\end{align}
This we can plug in into (\ref{fsym}) to get
\be
\bds{\omega}=\Big(\frac{1}{2}\dt\p^i \w \dt\p^j \Big(\frac{\partial}{\partial\partial_a \phi^j}\frac{\dt \tb{L}_a}{\dt \p^i}\Big)+ \dt\p^i \w \partial_b\dt\p^j \Big(\frac{\partial}{\partial\partial_a\partial_b \phi^j}\frac{\dt \tb{L}_a}{\dt \p^i}\Big) \Big) 
\ee
where \(\tb{L}_a=i_{\partial_a}\tb{L}\). The final expression for $\Omega_{BEF}$ therefore becomes
\be\label{BEF sym2}
\Omega_{BEF}=\int \tb{d}\sigma \w \Big(\frac{1}{2}\dt\p^i \w \dt\p^j \Big(\frac{\partial}{\partial\partial_a \phi^j}E^i_a\Big)+ \dt\p^i \w \partial_b\dt\p^j \Big(\frac{\partial}{\partial\partial_a\partial_b \phi^j}E^i_a\Big) \Big) 
\ee
where we have introduced the equation of motion $(d-1)$-forms $\bds E^i_a =E^i\bds\epsilon_a$. This result should be compared to \ref{BB2}: One can see that the term in parentheses is actually identical to \ref{BB2} and can be interpreted as the BEF-symplectic current $\omega_{BEF}$. The only difference is that the integral is taken over a generalized, "diffuse" time slice parametrized by the gradient of $\sigma$.   
Since this symplectic current obtained from the BEF-proposal matches with $\bds\omega^{BB}$, it will reproduce the Wald entropy formula for stationary black holes \cite{LeeWald1990,Wald:1993nt,Iyer:1994ys}. 
\\
\subsection{Relation between $\Omega_{BEF}$ and $\bds \omega^{BB}$ in general theories} 
The results from the previous section motivate the question about what happens if the equations of motion are not second order. In this section we will determine the precise relation between $\Omega_{BEF}$ and $\bds\omega^{BB}$ for arbitrary Langrangians.\\
Let us start again with the Lagrangian $d$-form  \(\bf{L}\) and the equation of motion $d$-forms \(\frac{\dt \tb{L}}{\dt\phi^i}=\bds E_i\). From \ref{BB IW relation} and \ref{LV} it is easy to check that
\be\label{gp1}
\bds{d}\bds{\omega}^{BB}=\dt\phi^i \w \dt \bds{E}_i
\ee
Now consider an operator \(i_R=R^i_{,I}\frac{\partial}{\partial \dt\phi^i_{,I}}\), where \(R^i\) are arbitrary local functions of coordinates, fields and their derivatives. We have
\be\label{gp2}
i_{R_2}i_{R_1}(\bds{d}\bds{\omega}^{BB})= i_{R_2}i_{R_1} (\dt\phi^i \w \dt \bds{E}_i).
\ee
We can use the fact that this inner product commutes with $\bds d$  and write
\be\label{gp3}
\bds{d} (i_{R_2}i_{R_1}\bds{\omega}^{BB})= i_{R_2}i_{R_1} (\dt\phi^i \w \dt \bds{E}_i)
\ee
Further, utilizing the definition \(\dt\bds{E}_i=\dt\phi^k_{,J}\frac{\partial}{\partial \phi^k_{,J}}\bds{E}_i\) we get
\be\label{gp4}
i_R(\dt\bds{E}_i)=R^k_{,J}\frac{\partial}{\partial \phi^k_{,J}}\bds{E}_i=:\delta_R \bds{E}_i.
\ee
Using \(i_R(\bds{\alpha} \wedge \bds{\beta})= (i_R \bds{\alpha}) \wedge \bds{\beta} + (-1)^{\deg \bds{\alpha}} \bds{\alpha} \wedge i_R \bds{\beta} \)  and the above equation, we get
\be\label{gp5}
R_1^i\dt_{R_2}\bds{E}_i-R_2^i\dt_{R_1}\bds{E}_i=\bds{d}\bds{\omega}^{BB}[R_1,R_2]
\ee
where we denote  \(\bds{\omega}^{BB}[R_1,R_2]=i_{R_2}i_{R_1}(\bds{\omega}^{BB})\). 
Now, we would like to write equation (\ref{gp5}) in terms of the linearized equation of motion operator \({\bds{Q}}_{ij}\)\footnote{\({\bds{Q}}_{ij}R^j=-\delta_R \bds{E}_i\)} defined in \ref{ft BEF 1}. Plugging this back into (\ref{gp5}), we get
\be\label{gp7}
\bds{d}\bds{\omega}^{BB}[R_1,R_2]=R_2^i {\bds{Q}}_{ij} R_1^j- R_1^i {\bds{Q}}_{ij} R_2^j.
\ee
Notice that the above formula is valid for any local functions \(R_1^i\) and \(R_2^j\). To make contact with $\Omega_{BEF}$  let us insert the sigmoid into this equation:
\be\label{gp8}
\bds{d}\bds{\omega}^{BB}[\sigma(x)R_1,R_2]=R_2^i {\bds{Q}}_{ij} (\sigma(x) R_1^j)-\sigma(x) R_1^i {\bds{Q}}_{ij} R_2^j
\ee
Furthermore,
\begin{multline}\label{gp9}
\frac{1}{2}(\bds{d}\bds{\omega}^{BB}[\sigma(x)R_1,R_2]+\bds{d}\bds{\omega}^{BB}[R_1,\sigma(x)R_2])-\sigma(x) \bds{d}\bds{\omega}^{BB}[R_1,R_2]\\= \frac{1}{2} R_2^i [ \bds{Q}_{ij},\sigma(x)]R_1^j- \frac{1}{2} R_1^i [ {\bds{Q}}_{ij},\sigma(x)]R_2^j.
\end{multline}
Next, we just "undo" the inner products with $R_1$ and $R_2$ and write the expression in terms of $\dt\p$. We get
\be\label{gp10}
-\frac{1}{2}\dt\phi^i [ {\bds{Q}}_{ij},\sigma(x)] \dt\phi^j= \frac{1}{2}\Big(\bds{d}\bds{\omega}^{BB}(\sigma(x)\dt\phi,\dt\phi)+ \bds{d}\bds{\omega}^{BB}(\dt\phi,\sigma(x)\dt\phi)\Big)-\sigma(x) \bds{d}\bds{\omega}^{BB}(\dt\phi,\dt\phi).
\ee
The right hand side is now nothing but $\Omega_{BEF}$, hence we obtain as a final result
 \be\label{gp11}
\Omega_{BEF}(\dt\phi,\dt\phi)=\int \frac{1}{2}\Big(\bds{d}\bds{\omega}^{BB}(\sigma(x)\dt\phi,\dt\phi)+ \bds{d}\bds{\omega}^{BB}(\dt\phi,\sigma(x)\dt\phi)\Big)-\sigma(x) \bds{d}\bds{\omega}^{BB}(\dt\phi,\dt\phi)
\ee
This relation is true  off-shell for any finite derivative theory and we will explicitly demonstrate it for a higher derivative scalar theory in section \ref{HD}.
\section{Hamilton function}
In this section we want to give a proposal\footnote{We got inspired by an expression for the Hamiltonian which was already given in \cite{Bernardes:2025uzg}  and later on used in calculations, see \cite{Bernardes:2025zkj}} how to obtain the Hamilton function associated to some vector field $\xi$  in spacetime. For any such vector field we can define the corresponding vector field on the configuration space \(\mathcal{F}\) as \(V_\xi=\mathcal{L}_\xi\phi \frac{\partial}{\partial\phi} \). Further, we can define an interior product with respect to \(V_\xi\) as \(i_{V_\xi}= (V_\xi)_{,I}\frac{\partial}{\partial \delta \phi_{,I}}\)\footnote{\(V_\xi\) is Grassmann even and \(i_{V_\xi}\) is Grassmann odd.}. Now, the Hamiltonian that generates the flow along the \(V_\xi\) should satisfy the relation
\be\label{Hm1}
\cancel{\delta} H_\xi=-i_{V_\xi}\Omega_{BEF}-\omega(\delta\phi,(\mathcal{L}_\xi \sigma)E(\phi))
\ee
where $E(\p)$ is the equation of motion. Here, we wrote the variation of $H_\xi$  as an \textsl{inexact differential} since we do not know yet if the right hand side is indeed $\dt$-exact.  Therefore, the idea is to compute \(i_{V_\xi}\Omega_{BEF}\) and massage it to obtain something  $\dt$-exact, probably up to a boundary term.
\\
Let us compute the RHS of (\ref{Hm1}):
\be\label{Hm2}
i_{V_\xi}\Omega_{BEF}=-\frac{1}{2}\omega(\mathcal{L}_\xi \phi,[Q_\phi,\sigma]\delta \phi)-\frac{1}{2}\omega(\delta\phi,[Q_\phi,\sigma]\mathcal{L}_\xi\phi).
\ee
It can be shown that both terms are equal using \(\tau\)-regulation:\footnote{Here, we are following the prescription of the section \ref{Sb11}.}.
\be\label{Hm3}
i_{V_\xi}\Omega_{BEF}=-\omega(\delta\phi,[Q_\phi,\sigma]\mathcal{L}_\xi\phi)
\ee
Further, we can open up the commutator and use 
\be
Q_\phi\mathcal{L}_\xi\phi=\sum_{n=1}^\infty \frac{1}{(n-1)!}L_n(\phi,..,\phi,\mathcal{L}_\xi\phi)=\sum_{n=1}^\infty \frac{1}{n!}\mathcal{L}_\xi L_n(\phi,..,\phi)
\ee
as well as
\begin{align}\label{Hm6}
\sigma Q_\phi\mathcal{L}_\xi\phi &=\sum_{n=1}^\infty \frac{1}{n!}\mathcal{L}_\xi (\sigma L_n(\phi,..,\phi))-\sum_{n=1}^\infty \frac{1}{n!}(\mathcal{L}_\xi \sigma) L_n(\phi,..,\phi)\\
 &=\sum_{n=1}^\infty \frac{1}{n!}\mathcal{L}_\xi (\sigma L_n(\phi,..,\phi))-(\mathcal{L}_\xi \sigma)E(\phi)
\end{align}
to get
\begin{multline}\label{Hm7}
i_{V_\xi}\Omega_{BEF}=\omega\Big(\delta\phi,\sum_{n=1}^\infty \frac{1}{n!}\Big[\mathcal{L}_\xi (\sigma L_n(\phi,..,\phi))-nL_n(\phi,..,\phi,\sigma\mathcal{L}_\xi\phi)\Big]\Big)\\-\omega(\delta\phi,(\mathcal{L}_\xi \sigma)E(\phi)).
\end{multline}
Comparing with \ref{Hm1} yields
\be\label{Hm8}
\cancel{\delta}H_\xi=-\omega\Big(\delta\phi,\sum_{n=1}^\infty \frac{1}{n!}\Big[\mathcal{L}_\xi (\sigma L_n(\phi,..,\phi))-nL_n(\phi,..,\phi,\sigma \mathcal{L}_\xi\phi)\Big]\Big),
\ee
which motivates us to define new brackets
\begin{align}\label{Hm9}
H_\xi^n(\phi_1,..,\phi_n)&=\mathcal{L}_\xi (\sigma L_n(\phi_1,..,\phi_n))-\sum_{i=1}^n L_n(\phi_1,..,\phi_{i-1},\sigma \mathcal{L}_\xi\phi_i,\phi_{i+1},..,\phi_n)
\end{align}
which give rise to 
\be\label{Hm10}
\cancel{\delta}H_\xi=-\omega\Big(\delta\phi,\sum_{n=1}^\infty \frac{1}{n!}H_\xi^n(\phi,..,\phi)\Big).
\ee
Now, for the Hamiltonian to exist we should be able to write the right hand side as a total variation. Let us consider the following quantity
\be\label{Hm11}
J_\xi=\omega\Big(\phi, \sum_{n=1}^\infty \frac{1}{(n+1)!}H^n_\xi(\phi,..,\phi)\Big)
\ee
and try to see whether its variation can produce (\ref{Hm10}):
\be\label{Hm12}
\delta J_\xi=-\omega \Big(\delta \phi, \sum_{n=1}^\infty \frac{1}{(n+1)!}H^n_\xi(\phi,..,\phi)\Big)-\sum_{n=1}^\infty \frac{1}{(n+1)!}\omega\Big( \phi, \delta H^n_\xi(\phi,..,\phi)\Big).
\ee
The first term is already in the required form, so we will massage the second term to make it look similar:
\begin{align*}
\omega\Big( \phi, \delta H^n_\xi(\phi,..,\phi)\Big)&= \omega\Big( \phi, \delta\Big[ \mathcal{L}_\xi (\sigma L_n(\phi,..,\phi))-nL_n(\phi,..,\phi,\sigma \mathcal{L}_\xi\phi)\Big]\Big)\\
&=\omega\Big( \phi, -n \mathcal{L}_\xi (\sigma L_n(\phi,..,\phi,\delta \phi))-n \delta L_n(\phi,..,\phi,\sigma \mathcal{L}_\xi\phi)\Big)\\&=-\omega\Big( \phi, n (\mathcal{L}_\xi \sigma) L_n(\phi,..,\phi,\delta \phi)+ \sigma n(n-1) L_n(\phi,..,\phi,\mathcal{L}_\xi\phi,\delta \phi) \\&+\sigma n L_n(\phi,..,\phi,\mathcal{L}_\xi\delta \phi) -n(n-1)  L_n(\phi,..,\phi, \delta \phi,\sigma \mathcal{L}_\xi\phi)\\&
-n L_n(\phi,..,  \phi,\sigma \mathcal{L}_\xi \delta\phi) \Big).\label{Hm14}
\end{align*}
Using the cyclicity of the \(L_n\) (and if necessary, $\tau$-regulation) and the antisymmetry of the bilinear form we get
\begin{multline}
\omega\Big( \phi, \delta H^n_\xi(\phi,..,\phi)\Big)=\omega\Big( \delta \phi, n  L_n(\phi,..,\phi,\mathcal{L}_\xi \sigma \phi)+  n(n-1) L_n(\phi,..,\mathcal{L}_\xi\phi,\sigma \phi) \\ -n(n-1)  L_n(\phi,..,  \phi,\sigma \mathcal{L}_\xi\phi)\
\Big)\\+\omega\Big(\mathcal{L}_\xi \delta \phi, n L_n(\phi,..,\sigma\phi)-\sigma n L_n(\phi,..,  \phi)\Big).
\end{multline}
This can be written in a compact way, using just the Leibniz rule for the Lie derivative:
\begin{multline}
\omega\Big( \phi, \delta H^n_\xi(\phi,..,\phi)\Big)=n\omega\Big( \delta \phi, \mathcal{L}_\xi (\sigma L_n(\phi,.., \phi))-nL_n(\phi,.., \phi,\sigma\mathcal{L}_\xi \phi)\Big)\\
+\mathcal{L}_\xi \omega(\delta \phi,n[L_n(\phi,..,\phi,\cdot ),\sigma]\phi )\\
=n\omega\Big( \delta \phi, H^n_\xi(\phi,..,\phi)\Big)+\mathcal{L}_\xi \omega(\delta \phi,n[L_n(\phi,..,\phi,\cdot ),\sigma]\phi ).\label{HM15}
\end{multline}
Plugging this equation back into (\ref{Hm12}), we get
\be\label{Hm16}
\delta J_\xi=-\omega \Big(\delta \phi, \sum_{n=1}^\infty \frac{1}{(n)!}H^n_\xi(\phi,..,\phi)\Big)-\sum_{n=1}^\infty \frac{n}{(n+1)!}\mathcal{L}_\xi \omega(\delta \phi,[L_n(\phi,..,\phi, \cdot),\sigma]\phi ).
\ee
It follows from the equation (\ref{Hm10}) that\footnote{Note that in the canonical setting the second term is just a total time derivative. So if we are interesting in the bulk Hamiltonian only, this term is irrelevant. }
\be\label{Hm17}
\cancel{\delta} H_\xi=\delta J_\xi-\mathcal{L}_\xi
\Theta_\sigma (\delta \phi,\phi)
\ee
where
\be\label{Hm18}
\Theta_\sigma (\delta \phi,\phi)= -\sum_{n=1}^\infty \frac{n}{(n+1)!} \omega(\delta \phi,[L_n(\phi,..,\phi,\cdot ),\sigma]\phi ).
\ee
\(\Theta_\sigma\) is nothing but the pre-symplectic potential in the BEF-formalism \footnote{Notice that substituting equation (\ref{Hm17}) into equation (\ref{Hm1}) yields an expression analogous to equation (29) of \cite{Hollands:2024vbe}.}. It can be checked using \(\tau\)- regulation that we indeed have 
\be
\dt\Theta_\sigma=\Omega_{BEF}.
\ee
For the Hamiltonian corresponding to the vector field \(\xi\) to exist, we therefore get the necessary condition\footnote{This condition may also be sufficient, assuming the absence of topological obstructions on the solution submanifold of the configuration space\cite{Wald:1999wa}.}
\be\label{Hm19}
\delta \cancel{\delta} H_\xi=0 ~\implies ~ \mathcal{L}_\xi \Omega_{BEF}=0.
\ee
A more complete and detailed study of conserved charges in this context definitely deserves further research, see also the recent work \cite{BEFcharges}.
\section{Examples}
In this section, we explicitly compute the BEF and BB symplectic forms, show the relation between them, and compute the associated Hamiltonian.
\subsection{Maxwell theory}
We now want to check our results for some explicit examples and the first one we consider is Maxwell theory. The calculation of $\Omega_{BEF}$  was already done in \cite{Bernardes:2025uzg}  so we just summarize the results. Starting from the action in $L_\infty$-form 
\be
S=-\frac{1}{2} \int d^dx \ A^\mu(\partial_\mu \partial_\nu-\Box \eta_{\mu\nu})A^\nu,
\ee 
the action for a small fluctuation around a given solution readily follows as 
\be
S_{fluc}=-\frac{1}{2} \int d^dx\  a^\mu(\partial_\mu \partial_\nu-\Box \eta_{\mu\nu})a^\nu,
\ee 
since Maxwell theory is a free theory. From a straightforward calculation it follows that
\be
\Omega_{BEF}=\int d^dx\  \partial_\mu \sigma\  \omega_{BEF}^\mu
\ee
with the presymplectic current
\be\label{presymcur Maxwell}
\omega_{BEF}^\mu=-\frac{1}{2}\Big( \dt A^\mu \wedge \partial_\nu \dt A^\nu + \dt A^\nu \wedge \partial_\nu \dt A^\mu-2\dt A^\nu \wedge \partial^\mu \dt A_\nu\Big).
\ee
From our previous analysis we expect this expression to be essentially equal to $\bds \omega^{BB}$  since Maxwell theory is second order in derivatives. So let us compute $\bds \omega^{BB}$  directly:\\
We can use equation \ref{BB2}, where only the first term will be non-zero in our case. We easily find  
\be
E_\mu=-(\partial_\mu \partial_\nu-\Box\eta_{\mu\nu})A^\nu
\ee
and in calculating the variation we just have to be careful to symmetrize over the indices $a$  and $b$  in \ref{BB2}. As a result, we get 
\be
\bds\omega^{BB}=-\frac{1}{2}\Big( \dt A^\mu \wedge \partial_\nu \dt A^\nu + \dt A^\nu \wedge \partial_\nu \dt A^\mu-2\dt A^\nu \wedge \partial^\mu \dt A_\nu\Big)\bds\epsilon_\mu
\ee
which precisely corresponds to \ref{presymcur Maxwell}.\\
It is instructive to use equation \ref{BB IW relation} and also calculate $\bds\omega^{IW}$  and the corner term $\bds B$. Starting from the Lagrangian top form 
\be
\bds L=-\frac{1}{2} A^\mu(\partial_\mu \partial_\nu-\Box \eta_{\mu\nu})A^\nu\bds\epsilon,
\ee 
we directly apply the operator $I^d_{\dt A^\rho}$ via equation \ref{ex homtopy1}. Noticing that the higher terms in the ellipsis vanish and, again, being careful about the symmetrization, we obtain
\be
I^d_{\dt A^\rho}\bds L=\frac{1}{4} \Big(\dt A^\mu \partial_\nu A^\nu -\partial_\nu\dt A^\mu  A^\nu+\dt A^\nu \partial_\nu A^\mu-\partial_\nu\dt A^\nu  A^\mu-2\dt A^\nu \partial^\mu A_\nu+2\partial^\mu \dt A^\nu A_\nu\Big)\bds\epsilon_\mu. 
\ee
This should be compared to the standard boundary term one gets via integration by parts. One sees that it actually makes a difference if one first moves out $\partial_\mu$ or $\partial_\nu$, although they commute. The only way to get a unique answer is to symmetrize over both possibilities and this is just what the Anderson operator does. To continue, let us apply $\dt$ to get 
\be
\bds \omega^{IW}=-\frac{1}{2}\Big( \dt A^\mu \wedge \partial_\nu \dt A^\nu + \dt A^\nu \wedge \partial_\nu \dt A^\mu-2\dt A^\nu \wedge \partial^\mu \dt A_\nu\Big)\bds\epsilon_\mu=\bds \omega^{BB}.
\ee
This means that for Maxwell theory in $L_\infty$-form, $\bds \omega^{IW}=\bds \omega^{BB}$ and the corner term $\bds B$  vanishes, which can also be shown directly. The improvement term 
\be
B_{\mu\nu}=-\frac{1}{2} \dt A_\mu \wedge \dt A_\nu 
\ee
which was found in \cite{Bernardes:2025uzg}  stems from the comparison to the presymplectic form for the standard action 
\be
S_{st}=-\frac{1}{4}\int d^dx F_{\mu\nu}F^{\mu\nu}
\ee
In general, rewriting the action in $L_\infty$-form will change  $\bds \omega^{IW}$ by a corner term, whileas  $\bds \omega^{BB}$ stays invariant since it only depends on the equations of motion.\\
Let us now calculate the Hamiltonian using the formula 
\be
J_\xi=\sum_{n=1}^\infty \frac{1}{(n+1)!} \omega (\p,H^n_\xi(\p,...,\p)).
\ee
For a free theory, there is only one term in the sum that explicitly reads 
\be
J_\xi=\frac{1}{2} \omega (\p,\mathcal{L}_\xi(\sigma Q\p)-Q(\sigma\mathcal{L}_\xi\p)).
\ee
We can use the Leibniz rule and straightforwardly manipulate the commutators to get 
\be\label{Ham com}
J_\xi=\frac{1}{2}\Big[\omega (\p,\mathcal{L}_\xi(\sigma) Q\p)-\omega(\p,[Q,\sigma]\mathcal{L}_\xi\p))- \omega (\p,\sigma [Q,\mathcal{L}_\xi]\p)\Big].
\ee
The last term actually vanishes if $\xi$  is constant because $Q$  does not have any field dependence. For Maxwell theory in flat background, we will just use $\xi=\partial_t$ and therefore omit this last term. Moreover, we will assume that $\sigma$  is a function of time only. Now, the first term is just the action with the additional localization factor $\partial_0\sigma$  included and reads in components\footnote{We use the mostly plus signature for $\eta_{\mu\nu}.$} 
\begin{align*}
  \int d^dx\ \partial_0\sigma\  A^\mu(QA)_\mu=\int d^dx\ \partial_0\sigma \Big(A_i\partial_i\pd_j A_j -A_i\pd_i\pd_0 A_0-A_0\partial_0\pd_i A_i \\ +A_0\partial_i\pd_iA_0+A_i\pd_0^2 A_i -A_i\partial_j\pd_j A_i\Big).  
\end{align*}
For the second term, we first need to compute 
\be
[Q,\sigma]_{\mu\nu}=\pd_\mu\pd_\nu\sigma-\eta_{\mu\nu}\Box \sigma+\pd_\mu\sigma\pd_\nu+\pd_\nu\sigma\pd_\mu-2\eta_{\mu\nu}\pd_\rho \sigma\pd^\rho.
\ee
Inserting and open up in components yields 
\begin{align*}
     &\int d^dx A^\mu[Q,\sigma]_{\mu\nu}A^\nu\\
  =&\int d^dx\ \partial_0\sigma \Big(-A_0\pd_i\pd_0 A_i-A_i\pd_i\pd_0 A_0+2A_i\pd_0^2 A_i\Big) +\pd_0^2 \sigma A_i\pd_0 A_i   \\
  =&\int d^dx\ \partial_0\sigma \Big(-A_0\pd_i\pd_0 A_i-A_i\pd_i\pd_0 A_0+A_i\pd_0^2 A_i+\pd_0 A_i\pd_0 A_i\Big) + \pd_0(A_i \pd_0 A_i \pd_0 \sigma) 
\end{align*}
where in the second step we used integration by parts to remove the second derivative of $\sigma$. Since we are interested in the bulk Hamiltonian, we set \(H=J_{\partial_t}\) and by adding up all together we obtain
\be
H=\frac{1}{2} \int d^dx\ \pd_0 \sigma \Big(A_0\partial_i\pd_iA_0+\pd_0 A_i\pd_0 A_i+A_i\partial_i\pd_j A_j-A_i\partial_j\pd_j A_i\Big)+\pd_0(A_i \pd_0 A_i \pd_0 \sigma).
\ee
To bring this into the desired form, we need partial integration for the spatial derivative as well:
\begin{align*}
H=\frac{1}{2} \int d^dx\ \pd_0 \sigma \Big(A_0\partial_i \pd_iA_0+\pd_0 A_i\pd_0 A_i-\partial_i A_i\pd_j A_j+\partial_j A_i\pd_j A_i\Big)\\
+\pd_0(A_i \pd_0 A_i \pd_0 \sigma)+\pd_i(\pd_0 \sigma A_j F_{ji}).   
\end{align*}
Substituting the expressions for the electric and magnetic field $E_i=F_{0i}$ and $B_i=\frac{1}{2}\epsilon_{ijk}F_{jk}$ yields
\begin{align*}
H=\frac{1}{2} \int d^dx\ \pd_0 \sigma \Big(E^2 +B^2 +A_0\partial_i \pd_iA_0-\partial_i A_0 \pd_iA_0+2\pd_i A_0 \pd_0 A_i\Big)\\
+\pd_0(A_i \pd_0 A_i \pd_0 \sigma)+\pd_i(\pd_0 \sigma A_j F_{ji}).   
\end{align*}
With one more partial integration in the last two terms of the first line this can be simplified to 
\begin{align*}
H=\frac{1}{2} \int d^dx\ \pd_0 \sigma \Big(E^2 +B^2 -2A_0\pd_i E_i\Big)
+\pd_0(A_i \pd_0 A_i \pd_0 \sigma)\\
+\pd_i(\pd_0 \sigma (A_j F_{ji}+A_0 F_{0i}+A_0\pd_0A_i)).   
\end{align*}
The first term is just the canonical Hamiltonian of Maxwell theory, so we have obtained the desired result up to boundary terms. Note that $A_0$ acts as a Lagrange multiplier, enforcing the first class constraint $\pd_iE_i=0.$ The temporal total derivative will vanish since the $\pd_0\sigma$ factor localizes on the Cauchy surface. The spatial total derivative will in general give a non-vanishing result, we leave a careful analysis for future work\footnote{That will require us to carefully specify the Gibbons-York term and also take into account the second term of equation \ref{Hm17}.}.

\subsection{Higher Derivative Scalar}\label{HD}
In this section we would like to consider the simplest higher derivative scalar theory\footnote{We are grateful to Vinicius Bernardes for suggesting this example.}  which was also analyzed in \cite{Harlow:2019yfa}. The Lagrangian of the theory is given by
\be\label{HD1}
L=\frac{1}{2}\phi(\Box-  \Box^2)\phi.
\ee
and its equation of motion is simply
\be\label{HD2}
(\Box-  \Box^2)\phi=0.
\ee
For the pre-symplectic potential we have to imply the Anderson operator again. After a careful calculation we get 
\begin{multline}
 \Theta^\mu= \frac{1}{2}(\phi \pd^\mu \dt\phi- \dt \phi \pd^\mu \phi)-\frac{1}{2}\Big(\phi \Box \pd^\mu \dt \phi-\dt\phi \pd^\mu \Box \phi+\\\frac{2}{3}(\pd^\mu \pd^\nu \phi \pd_\nu \dt\phi- \pd^\mu \pd^\nu\dt\phi \pd_\nu \phi)+\frac{1}{3}(\Box \phi \pd^\mu \dt\phi- \Box\dt\phi \pd^\mu \phi)\Big).   
\end{multline}\label{HD2}
Notice the numerical factors coming from symmetrization of inequivalent ways to integrate by parts. Straightforwardly we obtain
\begin{align}
    \omega^{\mu IW}&=\dt\Theta^\mu =\dt\phi\wedge \pd^\mu \dt\phi-\dt\phi\wedge \Box \pd^\mu \dt \phi-\frac{2}{3}\pd^\mu \pd^\nu \dt\phi\wedge \pd_\nu \dt\phi-\frac{1}{3}\Box \dt\phi\wedge \pd^\mu  \dt\phi\\&=\dt\phi\wedge \pd^\mu \dt\phi-\dt\phi\wedge \Box \pd^\mu \dt \phi+\pd^\mu \dt \p \wedge \Box\dt\p +\frac{2}{3}\pd^\nu(\pd_\nu \dt\phi \wedge \pd^\mu \dt\phi)
\end{align}
For $\omega^{BB}$ we apply $I^d_{\dt\p}$ directly on the equations of motion, $\dt\p\frac{\dt L}{\dt\p}$, which leads to precisely the same result, so we have $\omega^{BB}=\omega^{IW}$. 
\\
Now, let us compute the BEF symplectic form: Using its definition we have
\be\label{HD4}
\Omega= \frac{1}{2}\int d^dx \ \dt\phi \wedge [\Box- \Box^2, \sigma]\dt \phi.
\ee
We already know that \([\Box,\sigma]=\Box\sigma+2 \pd^\mu \sigma\pd_\mu \) and it can easily be checked that
\be\label{HD5}
[\Box^2,\sigma]=\Box^2 \sigma +4\Box \pd^\mu \sigma\pd_\mu +4\pd^\mu \pd^\nu \sigma \pd_\mu \pd_\nu+2 \Box\sigma \Box +4 \pd_\mu\sigma\Box \pd^\mu  
\ee
After inserting this into equation \ref{HD4} we simply get
\begin{multline}\label{HD6}
    \Omega_{BEF}=\int d^dx \  \dt\phi \wedge \pd^\mu\sigma \pd_\mu\dt\phi-\\\frac{1}{2} \int d^dx \ \dt\phi \wedge \Big[ 4\Box \pd^\mu\sigma\pd_\mu \dt\phi +4\pd^\mu \pd^\nu \sigma \pd_\mu \pd_\nu \dt\phi+2 \Box\sigma \Box \dt\phi+4\pd_\mu\sigma\Box \pd^\mu \dt\phi\Big]. 
\end{multline}
Now the idea is to use partial integration to shift the derivatives from the sigmoid and transform the above expression into something of the form \(\Omega=\int \pd_\mu \sigma ~\omega^\mu+\pd_\mu X^\mu \), i. e. terms where only one derivative acts on $\sigma$ plus a total derivative term. This procedure is not unique: For the first term in the brackets, it makes a difference if one first integrates by parts for $\pd_\mu$ or one of the $\pd_\nu$ in $\Box$. In the spirit of the Anderson homotopy operator we symmetrize over all options which yields

\begin{align}
\omega^\mu&=\dt\phi\wedge \pd^\mu\dt\phi-\dt\phi \wedge \Box \pd^\mu\dt\phi+\frac{2}{3}\pd_\nu \dt\phi \wedge \pd^\nu \pd^\mu \dt\phi+ \frac{1}{3} \pd^\mu \dt\phi\wedge \Box\dt\phi=\omega^{BB}
\end{align}

and
\begin{multline}\label{HD8}
X^\mu =-\frac{4}{3}\dt\phi \wedge \pd^\mu \pd^\nu\sigma \pd_\nu \dt\phi-\frac{2}{3}\dt\phi \wedge \Box\sigma \pd^\mu \dt\phi -\frac{2}{3} \dt\phi \wedge \pd_\nu \sigma \pd^\mu \pd^\nu \dt\phi-\frac{1}{3}\dt\phi \wedge \pd^\mu \sigma \Box \dt\phi.
\end{multline}
Hence we have reproduced  $\omega^\mu_{BB}$ in the bulk as expected. Now let us understand how the corner term $X^\mu$ arises and what its meaning is. We have derived the general formula
 \be\label{gp11}
\Omega_{BEF}(\dt\phi,\dt\phi)=\int \frac{1}{2}\Big(\bds{d}\bds{\omega}^{BB}(\sigma(x)\dt\phi,\dt\phi)+ \bds{d}\bds{\omega}^{BB}(\dt\phi,\sigma(x)\dt\phi)\Big)-\sigma(x) \bds{d}\bds{\omega}^{BB}(\dt\phi,\dt\phi).
\ee
Inserting our expression for $\omega^{BB}$ and rescaling the arguments with $\sigma(x)$ yields 
\begin{align}
    \omega^{\mu BB}(\sigma\dt\p,\dt\p)=\sigma\omega^{\mu BB}+\frac{2}{3}[\pd_\nu,\sigma]\dt\p\wedge\pd^\nu\pd^\mu\dt\p+\frac{1}{3}[\pd^\mu,\sigma]\dt\p\wedge\Box\dt\p,
\end{align}
\begin{align}
    \omega^{\mu BB}(\dt\p,\sigma\dt\p)=\sigma\omega^{\mu BB}-\dt\p\wedge[\Box\pd^\mu,\sigma]\dt\p+\frac{2}{3}\pd_\nu\dt\p\wedge[\pd^\nu\pd^\mu,\sigma]\dt\p+\frac{1}{3}\pd^\mu\dt\p\wedge[\Box,\sigma]\dt\p.
\end{align}
After plugging this into \ref{gp11}, we see that the $\pd_\mu\sigma\ \omega^{\mu BB}$-terms just give the bulk contribution to $\Omega_{BEF}$. This means, all the commutator terms together must sum up to the corner term $X^\mu$. One can straightforwardly show that this is indeed the case. From this calculation it should be clear why in higher derivative theories, $\Omega_{BEF}$ and $\Omega_{BB}$ in general differ by corner terms. Which of them will contribute in the end, depends on the choice of boundary conditions which are applied. This means, a priori, without knowing the boundary conditions, there is no way to determine a unique, canonical corner term.  
\\
 $X^\mu$ contains terms with one and two derivatives of $\sigma$.
 We conjecture that the higher derivative terms must be eliminated by boundary conditions. For instance, in the standard case where $\sigma=H(t-t_0)$, they would produce derivatives of $\dt(t-t_0)$, which only make sense in a distributional setting. Moreover, there is no natural geometrical object that can contract with those terms. If this conjecture is correct, $\Omega_{BEF}$ might be even more powerful than $\Omega_{BB}$, since it contains non-trivial information about applicable boundary conditions.  
 \\
Lastly, let us compute the Hamiltonian. The steps are precisely the same as for Maxwell theory, so we will not present the calculation in detail but just focus on the results. We start again with \ref{Ham com}  and set $\xi=\pd_t$, where $\sigma$ is a function of time only.\footnote{At this point it makes sense to switch to "mostly minus" signature to avoid minus signs in the time derivatives.} It yields
\begin{align*}
    H=\frac{1}{2}\int d^dx\Big[\pd_t\sigma\ \p(\Box-\Box^2)\p-\p[(\Box-\Box^2),\sigma]\pd_t\p\Big]\\=\frac{1}{2}\int d^dx\Big[\pd_t\sigma\ \p(\Box-\Box^2)\p-\pd_t^2\sigma\ \p\pd_t\p-2\pd_t\sigma\ \p\pd_t^2\p+\pd_t^4\sigma\ \p\pd_t\p+4\pd_t^3\sigma\ \p\pd_t^2\p\\+6\pd_t^2\sigma\ \p\pd_t^3\p+4\pd_t\sigma\ \p\pd_t^4\p-2\pd_t^2\sigma\ \p\pd_i\pd_i\pd_t\p-4\pd_t\sigma\ \p\pd_i\pd_i\pd_t^2\p\Big].
\end{align*}
Again, we need to remove the higher derivatives of $\sigma$ via partial integration. Since there are only time derivatives, there is no ambiguity and the procedure yields
\begin{align*}
   H=\frac{1}{2}\int \ d^dx\Big[\pd_t\sigma\ \p(\Box-\Box^2)\p+\pd_t\sigma\Big(-2\p\pd_t^2\p+(\pd_t \p)^2+\p\pd_t^2\p+(\pd_t^2\p)^2\\-2\pd_t\p\pd_t^3\p+\p\pd_t^4\p+2\pd_t\p\pd_i\pd_i\pd_t\p-2\p\pd_i\pd_i\pd_t^2\p\Big)+\pd_t \Big(-\pd_t\sigma\ \p\pd_t\p\\+\pd_t^3\sigma\ \p\pd_t\p+3\pd_t^2\sigma\ \p\pd_t^2\p-\pd_t^2\sigma(\pd_t\p)^2-\pd_t\sigma\ \pd_t^2\p\pd_t\p-2\pd_t\sigma\ \p\pd_i\pd_i\pd_t\p\Big)\Big].
\end{align*}
Splitting the first term in space and time derivatives and rearrange the terms gives the final result 
\begin{align*}
   H=\frac{1}{2}\int \ d^dx\  \pd_t\sigma\Big[ (\pd_t\p)^2+\pd_i\p\pd_i\p+(\pd_t^2\p)^2-2\pd_t\p\pd_t^3\p-2\pd_i\pd_t\p\pd_i\pd_t\p-(\pd_i\pd_i\p)^2\Big]\\+\pd_t\Big[-\pd_t\sigma\ \p\pd_t\p+\pd_t^3\sigma\ \p\pd_t\p+3\pd_t^2\sigma\ \p\pd_t^2\p-\pd_t^2\sigma(\pd_t\p)^2-\pd_t\sigma\ \pd_t^2\p\pd_t\p-2\pd_t\sigma\ \p\pd_i\pd_i\pd_t\p\Big]\\+\pd_i\Big[\pd_t\sigma(-\p\pd_i\p+\pd_t\p\pd_i\pd_t\p-\p\pd_i\pd_j\pd_j\p+\pd_i\p\pd_j\pd_j\p\Big].
\end{align*}
In the first two terms, we recognize the well-known Hamiltonian of a massless scalar. The interesting part though are the boundary terms: The temporal boundary term will not contribute since it localizes on the generalized Cauchy surface. The spatial boundary term however will contribute, depending on the boundary conditions which are applied. Observe that the first term, $-\p\pd_i\p$, is just the Gibbons-Hawking term which needs to be added to the action in case of Dirichlet boundary conditions, see \cite{Stettinger:2024uus}. We conjecture that although we have nowhere assumed the presence of a spatial boundary, the new formula yields non-trivial information about boundary terms. 
\subsection{Schrödinger theory}
As a last example we want to demonstrate our results for a non-covariant theory, so we consider free non-relativistic Schrödinger theory. We have no reason to expect a priori that our results will survive because all constructions we have used so far were fully covariant. Still, we find a perfect agreement, which suggests that $\Omega_{BEF}$ is consistent even for non-covariant theories. This can be understood from our previous analysis involving $\Omega_{BB}$. \\
The Lagrangian is given by 
\be
L=-\frac{1}{2}\Big(i\Psi \partial_t \Psi^\dagger- i\Psi^\dagger \partial_t \Psi-\frac{1}{2m}(\Psi \partial_i \partial^i \Psi^\dagger+\Psi^\dagger \partial_i \partial^i \Psi) \Big).
\ee
We will first of all calculate $\Omega_{BEF}$, so let us deduce $Q$: It is useful to use a two dimensional vector notation and rewrite $L$  as 
\be
L=-\frac{1}{2} \begin{pmatrix}
  \Psi &  \Psi^\dagger
\end{pmatrix}\begin{pmatrix}
0 & i\partial_t-\frac{1}{2m}\partial_i\partial^i\\
-i\partial_t-\frac{1}{2m}\partial_i\partial^i & 0
\end{pmatrix}
\begin{pmatrix}
  \Psi \\
  \Psi^\dagger
\end{pmatrix}
\ee
now $Q$ can be read off to be just the matrix in the middle. 
The commutator with $\sigma$  can be straightforwardly evaluated and yields 
\be
[Q,\sigma]=\begin{pmatrix}
 0 & i\partial_t\sigma-\frac{1}{2m}(\partial_i\partial^i\sigma+2\partial_i\sigma\partial^i)\\
-i\partial_t\sigma-\frac{1}{2m}(\partial_i\partial^i\sigma+2\partial_i\sigma\partial^i) & 0
\end{pmatrix}.
\ee
This gives us as a final result
\be\label{Omega Schrö}
\Omega_{BEF}=\int d^dx\   \partial_t \sigma  \ i\dt\Psi\wedge\dt\Psi^\dagger -\partial_i\sigma \frac{1}{2m}(\dt\Psi\wedge\partial^i \dt\Psi^\dagger+\dt\Psi^\dagger\wedge\partial^i \dt\Psi).
\ee
If $\sigma$  is a function of time only, it reduces to the standard result with the symplectic current 
\be
\omega=i\dt\Psi\wedge\dt\Psi^\dagger.
\ee
For a more general sigmoid function, the second term will become relevant\\
Let us compare this now to $\bds \omega^{BB}$: Since we have a second order theory again, we can apply equation \ref{BB2}. The two (dependent) equations of motion for $\Psi$  and $\Psi^\dagger$ are 
\be
E_\Psi=(i\partial_t+\frac{1}{2m}\partial_i \partial^i)\Psi,\ \ \ \  \ \ \ \ \ \ \ \ 
E_{\Psi^\dagger}=(-i\partial_t+\frac{1}{2m}\partial_i \partial^i)\Psi^\dagger
\ee
from which we get
\be
\bds \omega^{BB}= \ i\dt\Psi\wedge\dt\Psi^\dagger \bds\epsilon_t -\frac{1}{2m}(\dt\Psi\wedge\partial^i \dt\Psi^\dagger+\dt\Psi^\dagger\wedge\partial^i \dt\Psi)\bds\epsilon_i.
\ee
Here, $\bds\epsilon_t $ and $\bds\epsilon_i$  are the induced volume forms on a spacelike or timelike hypersurface, respectively. We can see that this expression precisely agrees with the symplectic current we can deduce from \ref{Omega Schrö}. Therefore, both the BEF proposal as well as the Barnich-Brandt symplectic current give the correct, although generalized result for a non-covariant theory.\\
Lastly, let us calculate the Hamiltonian: As in the previous examples, let us choose a sigmoid which is a function of time only, then the terms from \ref{Ham com}  become 
\begin{align}
 \omega (\p,\mathcal{L}_\xi(\sigma) Q\p)=\int d^dx\ \pd_t\sigma \Big(i\Psi \partial_t \Psi^\dagger- i\Psi^\dagger \partial_t \Psi-\frac{1}{2m}(\Psi \partial_i \partial^i \Psi^\dagger+\Psi^\dagger \partial_i \partial^i \Psi) \Big)  
\end{align}
and 
\begin{align}
\omega(\p,[Q,\sigma]\mathcal{L}_\xi\p)= \int d^dx\ \pd_t\sigma \Big(i\Psi \partial_t \Psi^\dagger- i\Psi^\dagger \partial_t \Psi\Big).   
\end{align}
The Hamiltonian is now easily found to be
\begin{align}
    H&=-\int d^dx\ \pd_t\sigma \ \frac{1}{4m}(\Psi \partial_i \partial^i \Psi^\dagger+\Psi^\dagger \partial_i \partial^i \Psi) \\ &=-\int d^dx\ \pd_t\sigma \ \Big(\frac{1}{2m}\Psi^\dagger \partial_i \partial^i \Psi+\frac{1}{4m}\pd_i(\Psi\pd^i\Psi^\dagger-\pd^i\Psi\ \Psi^\dagger)\Big)
\end{align}
With $p_i=i\pd_i$, the first term is just the ordinary Hamiltonian density $\Psi^\dagger\frac{p^2}{2m}\Psi$ which has been obtained up to a spatial boundary term. This boundary term is just the current density and hence again what we expect.

\section{ ~~Conclusion and outlook}
In this paper, we studied and analyzed the BEF proposal \cite{ Bernardes:2025uzg} from a Lagrangian viewpoint. We derived the BEF symplectic structure starting from an arbitrary \(L_\infty\)-Lagrangian within the covariant phase space formalism. The sigmoid function can be interpreted as effectively introducing a boundary by hand and provides a way to localize the degrees of freedom in non-local theories. 

Next, we established a precise relation between the BEF symplectic structure and the Barnich-Brandt symplectic form for general finite-derivative theories. In particular, we have shown that for theories with second-order equations of motion, the BEF symplectic structure coincides with the Barnich-Brandt symplectic form. This explains the emergence of the canonical corner term in general relativity within the BEF framework. 

The relation between the BEF and Barnich-Brandt constructions is not accidental: Both are defined via the equations of motion, and therefore invariant under ambiguities in the choice of Lagrangian and pre-symplectic potential. It is natural to interpret the BEF construction as an infinite-derivative completion of the Barnich-Brandt homotopy construction of an invariant symplectic form.

We also argue that the boundary contributions in the BEF symplectic form - specifically, those terms involving a single derivative of the sigmoid function that localize on the boundary - are potentially related to corner contributions, once appropriate boundary conditions are imposed.

Moreover, we suggest that the boundary terms in the BEF symplectic form involving higher derivatives of the sigmoid encode information about admissible boundary conditions. In particular, we argue that those terms must vanish after imposing the boundary conditions. This feature can be illustrated in a simple four-derivative example presented in the paper.

In addition, we systematically derive the Hamiltonian formulation for theories in \(L_\infty\)-description. The resulting Hamiltonian includes contributions from corner terms, as expected. Furthermore, we explicitly compute it for several examples to illustrate its compatibility with known results. 

Finally, using the example of the Schrödinger theory, we show that the BEF construction also applies to non-covariant Lagrangians. This can be understood from the fact that the BEF symplectic form is constructed using the equations of motion and can be expressed in terms of the Barnich-Brandt symplectic form for any finite-derivative theory. The Barnich-Brandt construction, which relies on Anderson's homotopy operator \cite{Anderson1992}, does not depend on covariance either.
\\
There are several interesting directions for future work:
\begin{itemize}
    \item  A more careful analysis of the boundary terms in the Hamiltonian would be worthwhile, as it requires a proper specification of boundary conditions and the inclusion of an appropriate Gibbons-York term. To achieve a more complete understanding of the Hamiltonian approach, it would also be interesting to construct the Peierls bracket or Poisson bracket within the BEF framework. This would help in understanding the charges and their algebra for theories in \(L_\infty\)-description, in particular for non-local theories, see also \cite{BEFcharges, Bernardes:2026rsk, Bernardes:2026kpx}.
    \item It would be interesting to understand black hole charges and the laws of black hole mechanics and thermodynamics \cite{Iyer:1994ys,Wald:1999wa,Wall:2015raa,Dong:2013qoa} in stringy theories using this framework.
    \item  It could as well play an important role in understanding entanglement entropy in string theory \cite{Balasubramanian:2018axm}. It may also shed light on how one can extend the notion of algebras of observables from local quantum field theory or perturbative quantum gravity \cite{Witten:2018zxz,Chandrasekaran:2022eqq,Ali:2023bmm,Ali:2024jkx} to string theory.
    \item Moreover, there is still no satisfactory way of implementing boundary conditions in non-local theories, like string field theory for instance. We hope that our results could be a first hint about the requirements that admissible boundary conditions should fulfill.\\
\end{itemize}

\textbf{Note added:}
There is some overlap in the discussion of the Hamiltonian with the upcoming work of Bernardes, Erler and F\i rat, \cite{BEFcharges}. We declare that all the results were obtained independently.

\section*{Acknowledgments}
We acknowledge the support of the Department of Atomic Energy, Government of India, under project no.~RTI4019. We thank Ashoke Sen for many helpful discussions and valuable suggestions; MA is also grateful to him for suggesting the BEF work. We are further grateful to Vinicius Bernardes, Theodore Erler, and Atakan Hilmi Fırat for their comments on the draft and for useful discussions. GS would like to thank Raphaela Wutte and Friedrich Schöller for useful comments.
\newpage
\begin{appendices}
 \section{The  variational bi-complex}\label{app1}
 \subsection{Generalities}
This is a short introduction to the variational bi-complex which is used heavily in the covariant phase space formalism. The main idea is to introduce a second complex based on the space $\mathcal{F}$ of field configurations apart from the familiar de-Rham complex generated by the exterior derivative $\bds d$  in spacetime\footnote{Spacetime differential forms will be denoted by bold faced letters.}. 
Consider a one-parameter family of field configurations $\phi(\lambda)$ in $\mathcal{F}$  and define 
\be \label{V}
\delta \p= \dl \p(\ld)\Big|_{\ld=0}.
\ee
Here, $\p\equiv \p(0)$ is our background configuration, hence $\dt\p$ depends on $\p$ . In fact, $\dt$ is just the exterior derivative on the Banach manifold $\mathcal{F}$  and generates differential forms on $\mathcal{F}$ in the usual manner. We call a differential form of $\bds d $-degree $p$ and $\dt$-degree $q$  simply a $(p,q)$-form. Note that while $p$ is bounded by the spacetime dimension, $q$ is in general unbounded since $\mathcal{F}$ is infinite-dimensional. 
Finally, let us summarize the different gradings involved: The original vector space $\mathcal{H}$ might already come with its own grading, for instance the ghost number in BRST-quantized field theories. In this paper and all the examples, we will however suppress this grading to not clutter the formulas with signs. Connected to that, there is the artificial grading we used to construct the vector space $V$, see \ref{def V}.  Moreover, there are the gradings $p$ and $q$ generated by $\bds d$  and $\dt$, respectively. Both  $\bds d$  and $\dt$ are Grassmann odd objects, consequently we have 
\be
\{\bds d, \dt\}=0
\ee
\\
 \subsection{Jet bundles}
 A jet bundle is the arena where variational calculus is naturally formulated. The idea is as follows: Start with a vector bundle over spacetime such that the space of smooth sections is just the space $\mathcal{F}$ of field configurations $\p(x)$. Now enlarge the fiber by taking the direct sum with all spacetime derivatives $\p_{,a}$ , $\p_{,ab}$ , ..., $\p_{,I}$ where $I$ is a multi-index\footnote{We assume that all partial derivatives commute such that the order of the indices inside $I$ is irrelevant. }. In this way, all the derivatives are a priori treated as independent quantities. At some point we just restrict ourselves to sections where the fiber coordinates $\p_{,I}$ indeed match with the corresponding derivative $\partial_I\p$. This independent treatment of derivatives is exactly what we use in the Euler-Lagrange equation.  
To see this, let us define a couple of useful operators: 
The total differential on $\mathcal{F}$  is given by
\be\label{vertical der}
\dt=\delta \p_{,I}\w\frac{\partial}{\partial\p_{,I}}
\ee
There are two types of summations here: The first over 
the multi-index \(I\) and the second, which we have 
suppressed from the beginning, over all fields present in the theory. \\
The generalized Euler operator is defined as 
\be\label{Gen Euler}
\frac{\dt}{\dt\p_{,I}}=\sum_{J}(-1)^{|J|}\binom{|I|+|J|}{|J|} \partial_J \frac{\partial}{\partial \p_{,IJ}} 
\ee
where \(|J|\) is just the count of the multi-index \(J\). It can be seen easily from the above formula that the action of  \(\frac{\dt}{\dt\p}\) (i. e. $I=\emptyset $ ) on the Lagrangian yields the equations of motion. \\
Finally let us introduce the so-called \textsl{Anderson homotopy operator} which is defined as
\be\label{homoptopy operator}
I^p_{\dt\p}\tb{T}=\sum_I \frac{|I|+1}{d-p+|I|+1}\partial_I\Big[ \dt\p \w \frac{\dt}{\dt\p_{I,b}}i_{\partial_b}\tb{T}\Big]
\ee
where \(\tb{T}\) is a \((p,q)\) form. Basically, $I^p_{\dt\p}$  maps $(p,q)$-forms to $(p-1,q+1)$-forms. Note it contains two sums over multi-indices: One explicit over $I$  and one implicit contained in $\frac{\dt}{\dt\p_{I,b}}$. On top forms and \((d-1)\)-forms on spacetime it acts as
\begin{align}
I^d_{\dt\p}(L\bds{\ep})&=\dt\p \w \frac{\partial L \bds{\ep}_a }{\partial \p_{,a}}-\dt\p\w \partial_b \frac{\partial L \bds{\ep}_a }{\partial \p_{,ab}}+ \dt \p_{,b}\w
\frac{\partial L \bds{\ep}_a }{\partial \p_{,ab}}+.... \label{ex homtopy1}\\
I^{d-1}_{\dt\p}(L^c\bds{\ep_c})&=\frac{1}{2}\dt\p \w \frac{\partial L^c \bds{\ep}_{ac} }{\partial \p_{,a}}-\frac{1}{3}\dt\p\w \partial_b \frac{\partial L^c \bds{\ep}_{ac} }{\partial \p_{,ab}}+\frac{2}{3} \dt \p_{,b}\w
\frac{\partial L^c \bds{\ep}_{ac} }{\partial \p_{,ab}}\\
&+\frac{1}{4}\dt\phi \wedge\partial_{,be}\frac{\partial L^c \bds{\ep}_{ac} }{\partial \p_{,abe}}-\frac{1}{2}\dt\phi_{,b} \wedge\partial_{,e}\frac{\partial L^c \bds{\ep}_{ac} }{\partial \p_{,abe}}+ \frac{3}{4}\dt\phi_{,be} \wedge\frac{\partial L^c \bds{\ep}_{ac} }{\partial \p_{,abe}} ...\label{ex homtopy2}
\end{align}
where \(\bds{\ep}\), \(\bds{\ep_a}\) and \(\bds{\ep_{ac}}\) are volume forms of codimension zero, one and two, respectively. As it should be clear from the above expression, the Anderson homotopy operator just implements integration by parts. This manifests itself as well through the two following properties:
\begin{align}
    \bds{\dt}&=\dt\p\frac{\dt}{\dt\p}-\tb{d}I^d_{\dt\p} && \text{when it acts on \(d\)-forms} \label{homtopy alg1}\\
    \bds{\dt}&=I^{p+1}_{\dt\p}\tb{d}-\tb{d}I^p_{\dt\p} && \text{when it acts on \(p\)-forms \((p<d)\)}\label{homtopy alg2}
\end{align}
The first one will be crucial to fix the ambiguity of the presymplectic potential $\bds\Theta$, see section (\ref{BBSF1}). Moreover, it can be straightforwardly shown that
\be
[\bds{\dt},I^d_{\dt\p}]=0.
\ee
\end{appendices}

\end{document}